\documentclass[10pt,aps,prb,notitlepage,twocolumn,superscriptaddress]{revtex4-2}

\usepackage{amsmath}
\usepackage{amssymb}
\usepackage{graphicx}
\usepackage{epstopdf}
\usepackage[colorlinks=true]{hyperref}
\usepackage{float}
\usepackage{multirow}
\usepackage{subcaption}
\usepackage{array}

\usepackage[export]{adjustbox}

\newcommand{\sgn}{\mathrm{sgn}}

\begin{document}
\title{Tuning the Chern number of Kitaev quantum spin liquid}

\date{\today}

\author{Seong Jun Kwon}
\affiliation{Department of Applied Physics, The University of Tokyo, Bunkyo, Tokyo 113-8656, Japan}
\affiliation{Department of Physics and Natural Science Research Institute, University of Seoul, Seoul 02504, Republic of Korea}
\author{Kyusung Hwang}
\email{kyusung.hwang@khu.ac.kr}
\affiliation{Department of Applied Physics, Kyung Hee University, Yongin 17104, Republic of Korea}
\author{Suk Bum Chung}
\email{sbchung0@uos.ac.kr}
\affiliation{Department of Physics and Natural Science Research Institute, University of Seoul, Seoul 02504, Republic of Korea}
\affiliation{School of Physics, Korea Institute for Advanced Study, Seoul 02455, Republic of Korea}

\begin{abstract}
It is now well understood that 
non-Kitaev spin interactions can be added to the Kitaev quantum spin liquid by applying external fields. Recent years have seen intensive discussion on the possible phase transitions that these spin interactions induce. In this paper, we will show through the perturbation theory the possibility of 
accessing a gapped spin liquid phase with a higher Chern number through, in contrast to the cases studied in literature, a continuous phase transition. Such a transition may be induced by external tuning parameters such as electric field and 
hydrostatic pressure.
\end{abstract}

\maketitle

\section{Introduction}

One of the most notable developments in 
quantum magnetism 
has been the proposal by Kitaev of the exactly soluble model of quantum spin liquid~\cite{Kitaev2006}.
The study of quantum spin liquids (QSLs)---exotic and elusive phases of matter that are electrically insulating and magnetically disordered yet involve many-body quantum entanglement~\cite{Savary2016, YZhou2017, Knolle2019, Broholm2020}---has repeatedly encountered difficulties even on the theoretical level in determining the stability of the quantum spin liquid ground state.
The Kitaev honeycomb model offers manifest clarity on this issue thanks to the exact solutions of the Kitaev spin liquid (KSL) ground state and its fractionalized quasiparticle excitations, Majorana fermions.
With the realization that transition metal atoms with strong spin-orbit coupling can approximate 
the Kitaev honeycomb model~\cite{Jackeli2009, Chaloupka2010}, the Kitaev magnetism became a subject of intensive experimental investigations as well~\cite{Winter2017, Takagi2019, Motome2020, Matsuda2025}.
Remarkably, experimental signatures of the KSL and Majorana fermions have been reported by the recent measurements of thermal Hall effect and heat capacity on the Kitaev material $\alpha$-RuCl$_3$~\cite{Kasahara2018, Yokoi2021, Tanaka2022, Imamura2024}.


Extensive literature on the stability of the KSL phase has arisen in recent years owing, on one hand, to the existence of candidate materials and, on the other hand, to the issue 
being much better defined on microscopic levels than 
any other quantum spin liquid models. 
In all candidate materials, physics of the transition metal atoms, 
essential for the bond-dependent spin interaction, 
never completely turns off the non-Kitaev spin interaction \cite{Jackeli2009, Chaloupka2010, Rau2014, Katukuri2014}, 
and these non-Kitaev interactions tend to favor magnetically ordered states over the KSL phase \cite{SKChoi2012, Rau2014, Sears2015, RDJohnson2015, HSKim2015, HBCao2016, Winter2016, JWang2019, JWang2020}. Furthermore, recent years have seen various studies on the additional {\it tunable} non-Kitaev spin interactions induced by an external electric field \cite{Bulaevskii2008, Miyahara2016, Bolens2018, Chari2021, Noh2024, Koyama2018} 
or 
hydrostatic pressure 
\cite{Chaloupka2015, Winter2016, Takikawa2019, Khomskii2020, HLiu2022, Wolf2022, XWang2023, Hauspurg2025, Rau2014a}.

That the non-Kitaev spin interactions 
can destabilize the KSL phase raises the possibility of its driving a continuous phase transition to topologically distinct quantum spin liquid phases. 
The 
transitions to a non-KSL phase driven by the non-Kitaev interactions investigated so far have been of the first-order, and 
all of them have been between the KSL and a spin-ordered and topologically trivial states, 
therefore involving both the change in topology and the breaking of spin rotational symmetry. 
On the other hand, starting with the Haldane honeycomb model \cite{Haldane1988}, 
it has been 
recognized that a continuous phase transition is possible when change in topology occurs without any symmetry breaking. 
Investigations of continuous phase transitions in the Kitaev materials have so far 
concentrated mainly on 
the transition between the KSL phases with the opposite chirality \cite{Kitaev2006, Hwang2022, Noh2024}. 
Meanwhile, recent variational Monte Carlo calculations \cite{JWang2019, JWang2020} have found a topologically non-trivial quantum spin liquid phase with a higher winding number (a `proximate KSL') stabilized by sufficiently strong non-Kitaev spin interactions, 
but the direct transition from the KSL to the proximate KSL 
have been found to be of the first order, {\it i.e.} 
discontinuous. A proximate KSL topologically distinct from the KSL yet with a continuous transition from KSL would be more convenient to identify experimentally, 
especially if this continuous transition can be accessed through external tuning parameters. 
While one type of non-Kitaev spin interaction was shown to induce a continuous transition from the KSL to a proximate KSL state \cite{Fujimoto2020}, it has not been accompanied by discussion on symmetry-allowed tuning of 
non-Kitaev interaction terms. 

In this paper, we investigate possible continuous transitions from the KSL to proximate KSLs by the non-Kitaev spin interactions that are tunable with external electric and magnetic fields and hydrostatic pressure.
By performing a perturbative analysis,
we find that with a weak but finite magnetic field along the (111) direction, the increasing electric field along the (111) direction or hydrostatic pressure can lead to the continuous topological transition from the non-Abelian KSL with the Chern number of $\pm 1$ to the proximate Abelian KSL with the Chern number of $\mp 2$~\cite{Kitaev2006, JWang2019, JWang2020, Fujimoto2020, SSZhang2020}.
At zero magnetic field, we identify a variety of interesting nodal structures in the excitation spectrum including (i) multiple Majorana cones, (ii) Majorana quadratic band touching, and (iii) Majorana line node.
We analyze how magnetic fields gap out the zero-field nodal structures, leading to the KSLs with nonzero Chern numbers.
Interestingly, certain types of nodal structures survive under magnetic fields in the form of gap closing at the topological transition
between the distinct KSLs with different Chern numbers. 
We remark on implications of our results in quantum information applications of the KSL~\cite{YLiu2022, Klocke2024}.

The paper is organized as follows.
In Sec.~\ref{sec:model}, we present our model Hamiltonian and discuss its symmetry with and without external tuning; this discussion will make clear how some of the non-Kitaev interactions can be externally tuned. 
By requiring our external tuning parameters, {\it e.g.} electric field, magnetic field, and hydrostatic pressure, to all respect the C$_3$ rotational symmetry around the (111) direction, 
our analysis is simplified, 
with the number of independent spin interaction parameters restricted. 
In Sec.~\ref{sec:perturbation}, we present our perturbative analysis on the non-Kitaev terms in the zero-flux sector, focusing on how effective longer-range Majorana hoppings arise along with their impacts on the topological structure of the KSL and the resulting topological transitions.
Lastly, we present discussion our results, including implications on quantum information applications of the KSL in Sec.~\ref{sec:discussion}.
Details of the perturbation theory and the topological transitions are provided in Appendices.

\begin{figure}[tb]
    \centering
    \includegraphics[width=\linewidth]{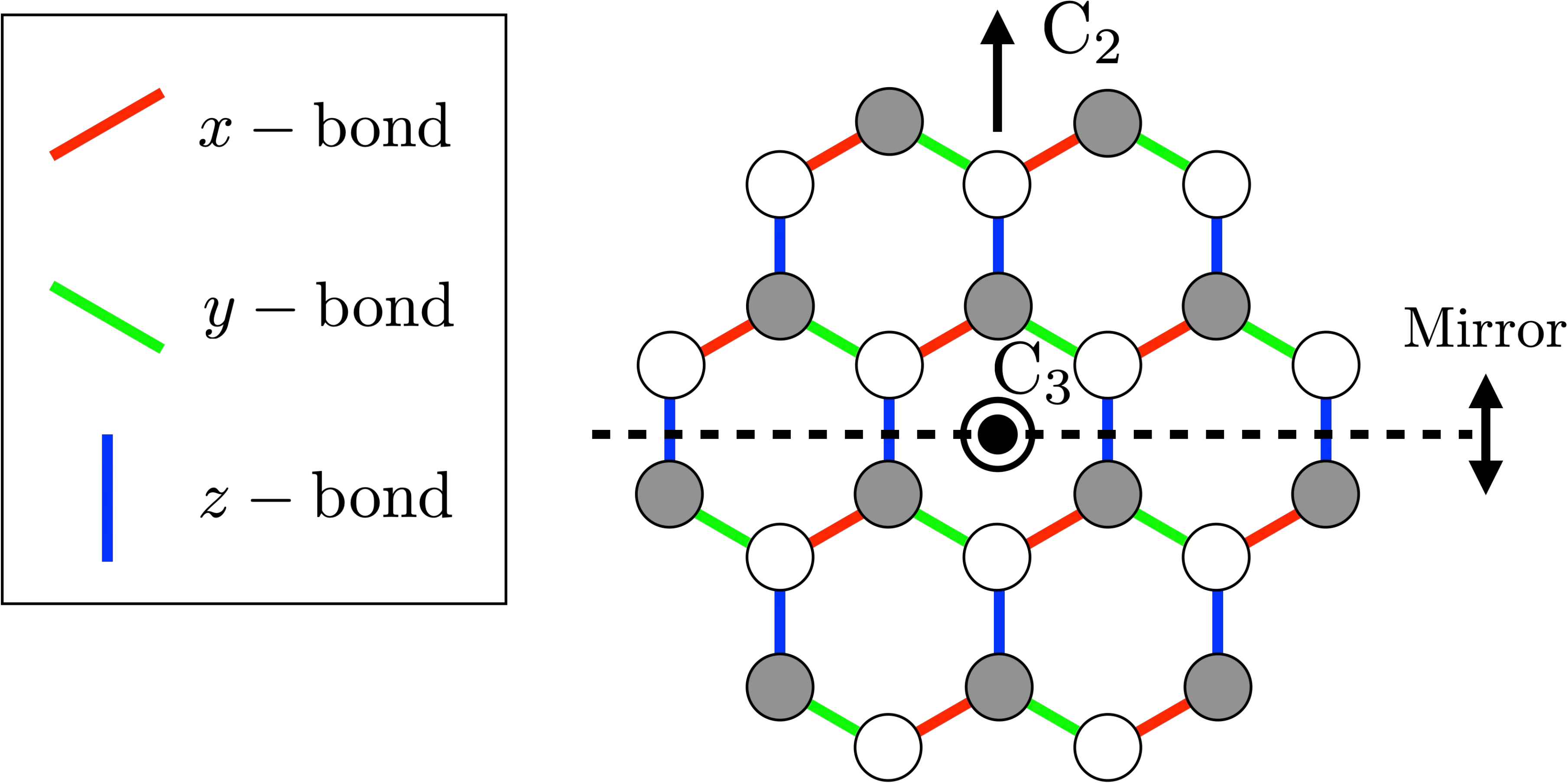}
    \caption{Lattice symmetries of the Kitaev Honeycomb model. Three types of bonds of the Kitaev model are denoted by three different colors: $x$-bond (red), $y$-bond (green), and $z$-bond (blue). $A$ and $B$ sublattices of the honeycomb lattice are indicated by gray and white circles, respectively. The Kitaev model has the C$_3$ rotation symmetry about the normal axis to each hexagon plaquette, the C$_2$ rotation symmetry about the axis of each bond, and the mirror reflection symmetry with respect to the normal plane bisecting each bond (dashed line).}
    \label{fig:lattice-symm}
\end{figure}

\section{Model Hamiltonian and symmetry\label{sec:model}}

We consider a generalized Kitaev model which contains various non-Kitaev interactions including Dzyaloshinskii-Moriya (DM) and $\varGamma'$ interactions and Zeeman coupling. Our model Hamiltonian takes the form,
\begin{equation}
H 
=
\sum_{\langle jk \rangle_\gamma} {\boldsymbol \sigma}_j \cdot {\bf M}_\gamma \cdot {\boldsymbol \sigma}_k 
- {\bf h} \cdot \sum_j {\boldsymbol \sigma}_j ,
\label{eq:model}
\end{equation}
where ${\boldsymbol \sigma}_j = (\sigma_j^x,\sigma_j^y,\sigma_j^z)$ is the three-component vector of the Pauli matrices operating on the spin-1/2 at site $j$. The $3\times3$ matrix ${\bf M}_\gamma$ describes exchange interactions of nearest-neighboring spins at $\gamma$-bonds, where $\gamma\in\{x,y,z\}$ (Fig.~\ref{fig:lattice-symm}).
To be specific, the matrix at $z$-bonds is given by
\begin{equation}
{\bf M}_z
=
{\bf M}_K + {\bf M}_{\Gamma^{\prime}} + {\bf M}_D
,
\label{eq:M_z}
\end{equation}
\begin{subequations}
\begin{eqnarray*}
{\bf M}_K & = & \left( \begin{array}{ccc}
0 & 0 & 0
\\
0 & 0 & 0
\\
0 & 0 & K
\end{array}
\right) \\
{\bf M}_{\Gamma^{\prime}} & = & \left(
\begin{array}{ccc}
0 & 0 & \varGamma'
\\
0 & 0 & \varGamma'
\\
\varGamma' & \varGamma' & 0
\end{array}
\right) \\
{\bf M}_{D} & = & \left(
\begin{array}{ccc}
0 & D_2 & -D_1
\\
-D_2 & 0 & D_1
\\
D_1 & -D_1 & 0
\end{array}
\right)
\end{eqnarray*}
\end{subequations}
{\it i.e.}, Kitaev, $\varGamma'$, and DM interactions.
The other matrices ${\bf M}_{x,y}$ for $x,y$-bonds can be obtained by cyclic permutations of ${\bf M}_{z}$ (equivalently C$_3$ rotations of the $z$-bond interaction term).
In a vector representation, the DM interactions can be represented by
\begin{equation}
{\bf D}_\gamma \cdot \left(\boldsymbol{\sigma}_j \times \boldsymbol{\sigma}_k\right),
\end{equation}
with the bond-dependent DM vectors,
\begin{eqnarray}
{\bf D}_x&=&(D_2,D_1,D_1)~~~~~{\rm for}~x{\rm -bond},
\\
{\bf D}_y&=&(D_1,D_2,D_1)~~~~~{\rm for}~y{\rm -bond},
\\
{\bf D}_z&=&(D_1,D_1,D_2)~~~~~{\rm for}~z{\rm -bond}.
\end{eqnarray}
In this representation, we choose the convention that site $j$ is on the $A$ sublattice and site $k$ is on the $B$ sublattice. 
The last term of $H$ represents the Zeeman coupling with a magnetic field along the out-of-plane direction,
\begin{equation}
{\bf h}=\frac{h}{\sqrt{3}}(1,1,1) .
\label{eq:B-field}
\end{equation}
Among the symmetries of the pure Kitaev model, our model preserves the C$_3$ rotational symmetry around the normal axis, {\it i.e.} the (111) direction, to each plaquette (see Table~\ref{tab:symm}).
%
%
%
%

\begin{table}[tb]
\begin{ruledtabular}
\begin{tabular}{lcccc}
Symmetry & $K$ & $\varGamma'$ & $D_{1,2}$ & ${\bf h}\parallel (111)$
\\
\hline
C$_3$~rotation & \checkmark & \checkmark & \checkmark & \checkmark
\\
C$_2$~rotation  & \checkmark & \checkmark & &
\\
Time~reversal  & \checkmark & \checkmark & \checkmark &
\\
Spatial~inversion  & \checkmark & \checkmark & & \checkmark
\\
Mirror~reflection & \checkmark & \checkmark & \checkmark &
\end{tabular}
\end{ruledtabular}
\caption{Preserved symmetries by each interaction of the model Hamiltonian $H$ [Eq.~(\ref{eq:model})]. See Fig.~\ref{fig:lattice-symm} for the definitions of the C$_3$ rotation, C$_2$ rotation, and Mirror reflection.}
\label{tab:symm}
\end{table}

Table~\ref{tab:symm} also makes it clear that, 
among the cases where the hydrostatic pressure, the applied external electric field, and the applied magnetic field are all present, 
our model possesses 
the highest symmetry possible. 
To begin with, the $\varGamma'$ terms represent the simplest spin interaction arising from the 
trigonal distortion for the spin-1/2 model on the honeycomb lattice with the edge-sharing octahedron structure; it preserves not only the C$_3$ rotational symmetry but all the symmetries of the Kitaev terms. 
It has been reported \cite{Winter2016, Bolens2018, Chaloupka2015, Khomskii2020, Hwang2022} that such trigonal distortion term can be induced by applying hydrostatic pressure. 
Secondly, while any non-zero DM interactions would break both the C$_2$ rotational symmetry around each bond axis and the spatial inversion exchanging the two sublattices of the honeycomb lattice, our model does preserve 
the C$_3$ rotational symmetry. 
Such DM interaction can be induced by applying a uniform electric field along the (111) direction, in which case 
there remains 
the 
mirror symmetry with respect to the planes perpendicularly bisecting 
each bond. 
This mirror symmetry leaves us with the two independent coefficients for the DM interactions ($D_1$ and $D_2$) as shown in Eq.~(\ref{eq:M_z}), each proportional to the magnitude of the applied electric field \cite{Rau2014, Bolens2018, Chari2021, Noh2024}. 
Lastly, any non-zero Zeeman field ${\bf h}$ will break the time-reversal symmetry preserved by all the spin interactions represented by the first term of Eq.~\eqref{eq:model}, but ${\bf h} \parallel (111)$ leaves the C$_3$ rotational symmetry intact. The discussion on how the DM and $\varGamma^\prime$ spin interactions 
arise from combination of electron hoppings and electronic interactions is presented in Appendix A.

\section{Perturbation analysis and topological transition\label{sec:perturbation}}

We investigate influence of the non-Kitaev interactions on the Kitaev spin liquid  state of the pure Kitaev model via a perturbation theory. To this end, the Hamiltonian is separated into two parts:
\begin{equation}
H=H_0+H_1,
\end{equation}
where $H_0$ is the pure Kitaev model 
$$
H_0 \equiv K \sum_{\langle jk \rangle_\gamma} \sigma_j^\gamma \sigma_k^\gamma
$$
and $H_1$ contains the non-Kitaev interactions ($D_1,D_2,\varGamma',h$) which are going to be considered as perturbations.
By performing a perturbative analysis with the non-Kitaev term, we find effective multi-spin interactions that lead to topological transitions in the Kitaev spin liquid.

\begin{table}[tb]
\begin{ruledtabular}
\begin{tabular}{lcl}
    {\rm Zeeman-type} & Every site & $\sigma_j^x$, $\sigma_j^y$, $\sigma_j^z$ \\
    \hline
    {$D_2$-{\rm type}} & $x$-bond & $\sigma_j^y \sigma_k^z$, $\sigma_j^z \sigma_k^y$ \\
    & $y$-bond & $\sigma_j^z \sigma_k^x$, $\sigma_j^x \sigma_k^z$ \\
    & $z$-bond & $\sigma_j^x \sigma_k^y$, $\sigma_j^y \sigma_k^x$ \\
    \hline
    {$D_1/\varGamma^{\prime}$-{\rm type}} & $x$-bond & $\sigma_j^x \sigma_k^y$, $\sigma_j^y \sigma_k^x$, $\sigma_j^x \sigma_k^z$, $\sigma_j^z \sigma_k^x$ \\
    & $y$-bond & $\sigma_j^y \sigma_k^z$, $\sigma_j^z \sigma_k^y$, $\sigma_j^y \sigma_k^x$, $\sigma_j^x \sigma_k^y$ \\
    & $z$-bond & $\sigma_j^z \sigma_k^x$, $\sigma_j^x \sigma_k^z$, $\sigma_j^z \sigma_k^y$, $\sigma_j^y \sigma_k^z$ \\
\end{tabular}
\caption{Classification of the non-Kitaev terms based on flux-creation patterns.}
\label{tab:classification-non-Kitaev}
\end{ruledtabular}
\end{table}

\subsection{Perturbation Theory of the Extended Kitaev Model}

\begin{figure}[tb]
\centering
\includegraphics[width=\linewidth]{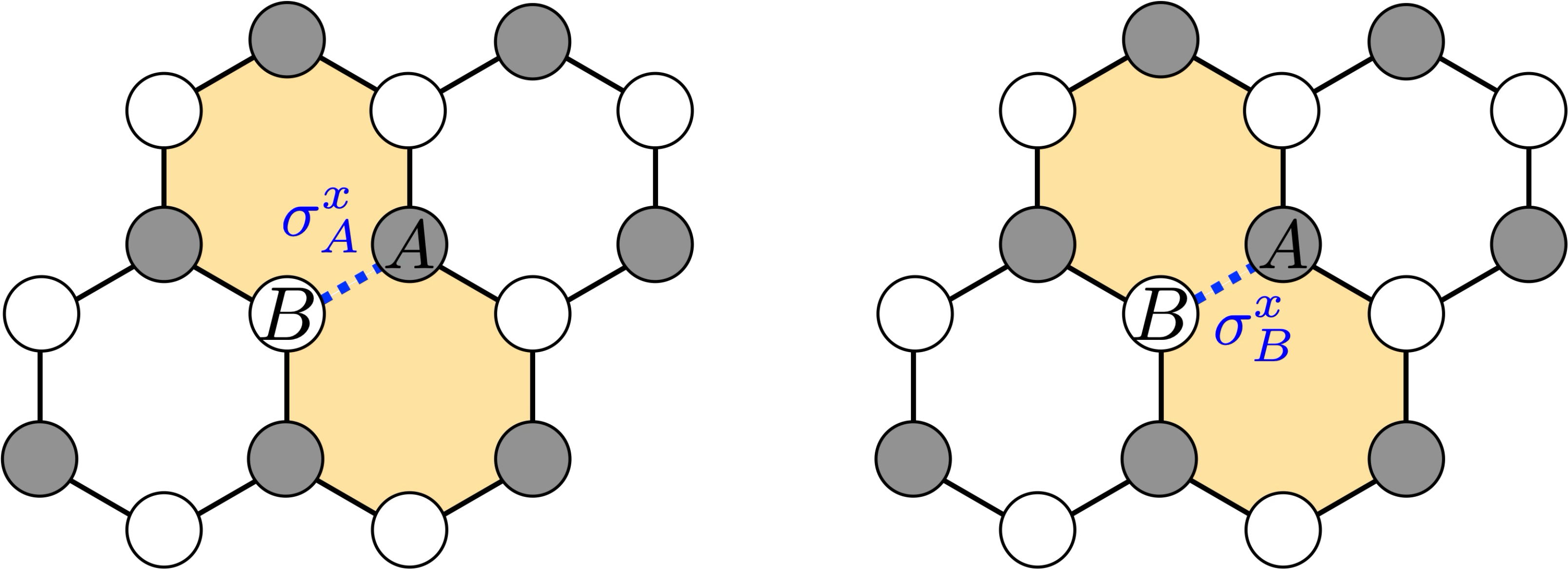}
    \caption{Flux creation by Zeeman-type terms at $x$-bond: $\sigma_{A}^x$ (left) and $\sigma_{B}^x$ (right). In each case, yellow plaquettes indicate $\mathbb{Z}_2$ fluxes created by the spin operator.
    The blue dashed line marks the sign-changed $\mathbb{Z}_2$ gauge field by the spin operator ($i b^x_A b^x_B:~+1 \rightleftarrows -1$).}
    \label{fig:Z-type}
\end{figure}

\begin{figure}[tb]
    \centering
    \includegraphics[width=\linewidth]{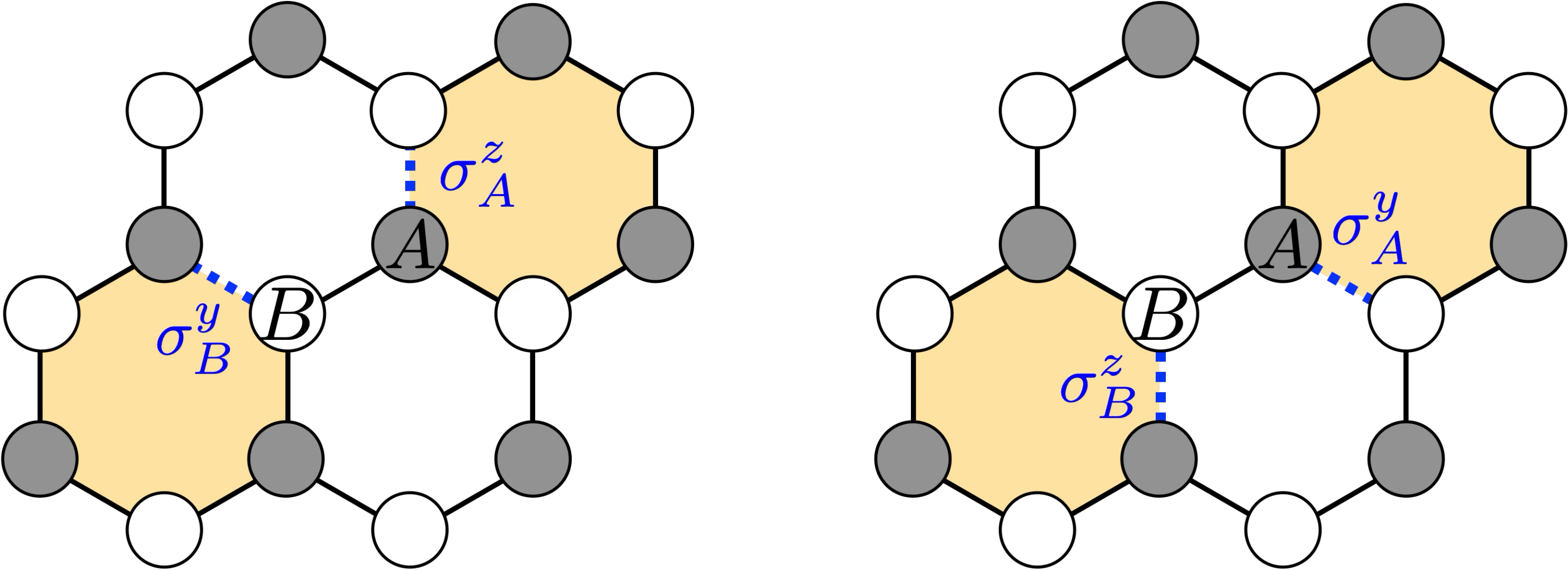}
    \caption{Flux creation by $D_2$-type terms at $x$-bond: $\sigma_{A}^z \sigma_{B}^y$ (left) and $\sigma_{A}^y \sigma_{B}^z$ (right). The blue dashed line again marks the sign-changed $\mathbb{Z}_2$ gauge field.}
    \label{fig:G-type}
\end{figure}

\begin{figure}[tb]
     \centering
    \includegraphics[width=\linewidth]{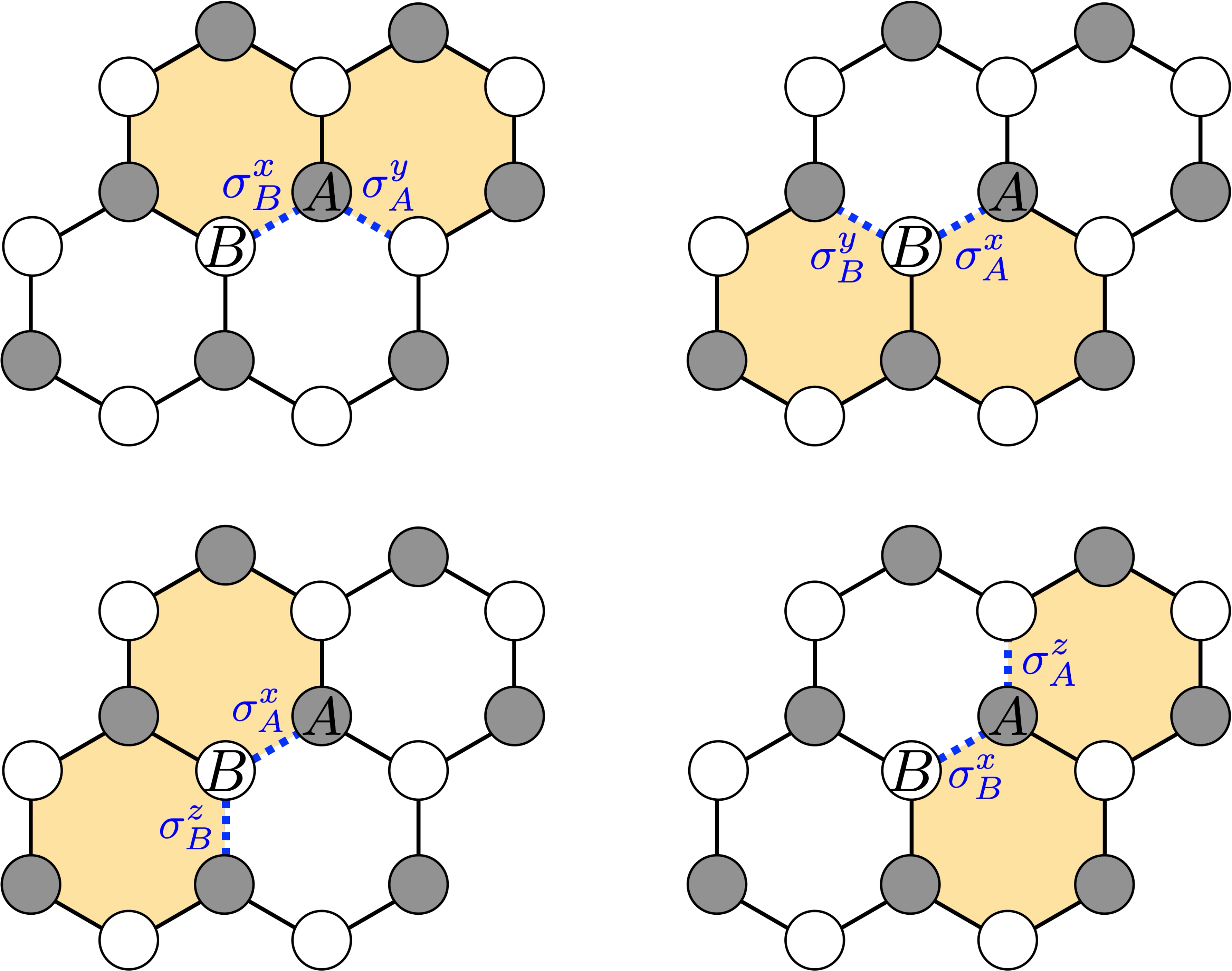}
        \caption{Flux creation by $D_1/\varGamma'$-type terms at $x$-bond: $\sigma_{A}^y \sigma_{B}^x$ (top left), $\sigma_{A}^x \sigma_{B}^y$ (top right), $\sigma_{A}^x \sigma_{B}^z$ (bottom left), and $\sigma_{A}^z \sigma_{B}^x$ (bottom right).}
        \label{fig:P-type}
\end{figure}

Discussion of the pure Kitaev model ($H_0$) is a prerequisite to present any perturbation theory. Two properties of $H_0$
are of particular interest. One is its commutation with the flux operator of any unit hexagon plaquette $p$ defined as \cite{Kitaev2006}
\begin{equation}
W_p = \prod_{\langle jk \rangle_\gamma \in p} \sigma_j^\gamma \sigma_k^\gamma;
\end{equation}
note that the plaquette flux operator also commutes with each other, $[W_p,W_{p'}]=0$. The other is that through a Majorana representation of spin-1/2 operators, the most symmetric form being \cite{Kitaev2006}
\begin{equation}
\sigma^\gamma_j = ib^\gamma_j c_j,
\label{EQ:spinMajorana}
\end{equation}
where $b^\gamma_j$ and $c_j$ are Majorana fermions, it can be written as the Hamiltonian 
where the nearest-neighbor hopping of the $c$-Majorana fermions is coupled to a $\mathbb{Z}_2$ gauge field,
\begin{equation}
H_0 = -iK\sum_{\langle jk \rangle_\gamma} (i b^\gamma_j b^\gamma_k) c_j c_k.
\end{equation}
These two properties are intimately related as $W_p$ turns out to be nothing more than the $\mathbb{Z}_2$ gauge flux, {\it i.e.}
\begin{equation}
W_p = \prod_{\langle A,B \rangle_\gamma \in p} (i b^\gamma_A b^\gamma_B),
\label{EQ:fluxMajorana}
\end{equation}
where A, B are the sublattice labels as shown in Figs.~\ref{fig:Z-type}, \ref{fig:G-type}, and \ref{fig:P-type}. Lastly, it has been shown that the $H_0$ ground state is in the zero flux sector, {\it i.e.} $W_p = +1$ for all plaquettes.

As our analysis perturbatively treats $H_1$ to obtain the effective Hamiltonian in the zero-flux sector, a natural starting point would be to study how the terms of $H_1$ affect the plaquette flux $W_p$. 
As we shall show here, no 
term in the $H_1$ 
leaves $W_p$ invariant. 
The underlying reason is that $\sigma_j^{\gamma}$ induces a $\mathbb{Z}_2$-flux pair in adjacent hexagons with a $\gamma$-bond coming in between, as depicted in Fig.~\ref{fig:Z-type} \cite{Kitaev2006, Hwang2022}; this can be seen from Eqs.~\eqref{EQ:spinMajorana} and \eqref{EQ:fluxMajorana}. It should be obvious that this is directly applicable to the linear spin terms  of the Zeeman Hamiltonian $H_{\rm Z}=-{\bf h}\cdot \sum_j {\bf \sigma}_j$. 
But the quadratic spin terms of $H_1$ also generate 
$\mathbb{Z}_2$-flux pairs, albeit of different types; the quadratic terms are classified into two types, $D_2$-type and $D_1/\varGamma^{\prime}$-type (Table~\ref{tab:classification-non-Kitaev}).
Flux patterns created by these quadratic terms are illustrated in Figs.~\ref{fig:G-type},~\ref{fig:P-type}
\vspace{10pt}

\begin{figure}[tb]
\centering
\includegraphics[width=\linewidth]{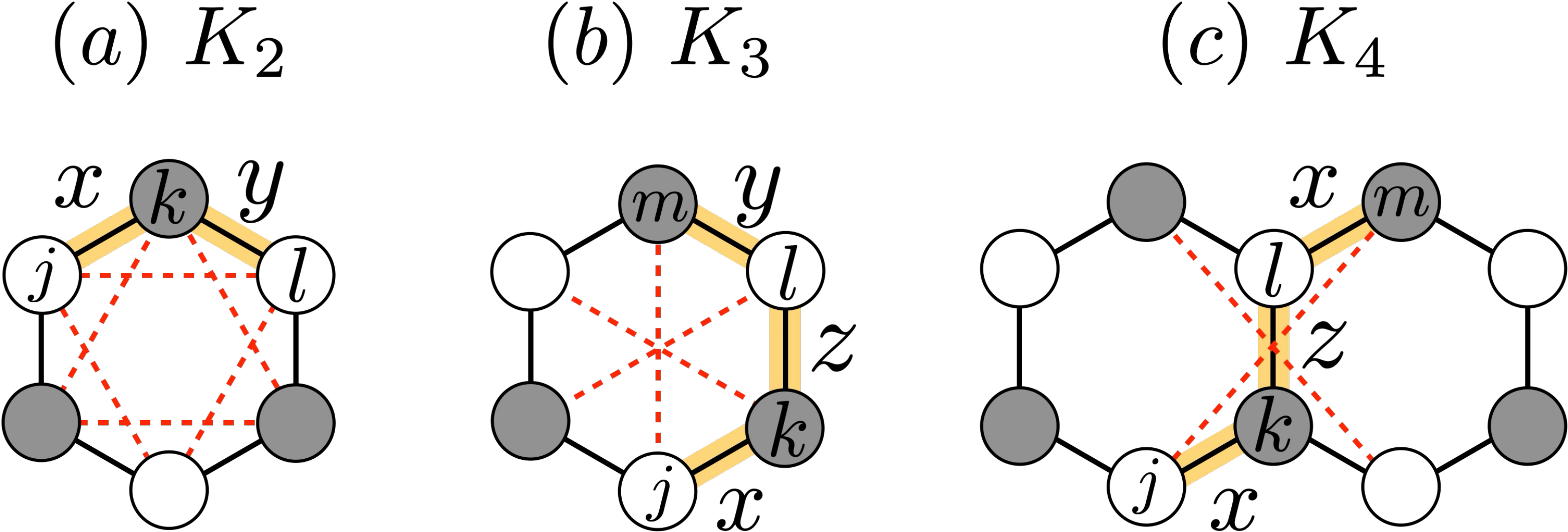}
\caption{
Illustration of the effective multi-spin interactions of $H_{\rm eff}$ [Eq.~(\ref{EQ:perturbSpin})].
(a) $K_2$ term on two consecutive Kitaev bonds $\langle jkl\rangle_{xy}$.
(b) $K_3$ term on three consecutive Kitaev bonds $\langle jklm\rangle_{xzy}$.
(c) $K_4$ term on three consecutive Kitaev bonds $\langle jklm\rangle_{xzx}$.
The red dashed lines depict the additional hopping paths of Majorana fermions resulting from the multi-spin interactions.
}
\label{fig:multi-spin-int}
\end{figure}

It can be seen clearly now that in the zero flux sector, all the first-order perturbation terms should vanish and the second-order terms are the lowest order terms after $H_0$, resulting in the effective zero-flux sector Hamiltonian of 
\cite{Kitaev2006, Takikawa2019, Fujimoto2020, Chari2021, Hwang2022, Noh2024}.
\begin{equation}
H_{\rm eff} = P_0 \left(H_0 - \sum_m \frac{H_1|m\rangle\langle m|H_1}{\Delta_m} \cdots\right) P_0  ,
\label{EQ:perturb}    
\end{equation} 
where $|m\rangle$ is the intermediate state with a  $\mathbb{Z}_2$-flux pair, $\Delta_m$ is the energy cost for creating the $\mathbb{Z}_2$-flux pair, and $P_0$ is the projection operator that projects to the zero-flux sector:
\begin{equation}
P_0=\prod_p\frac{1+W_p}{2}.
\end{equation}
we have established above that there are actually only two different values for $\Delta_m$'s, $\Delta \approx 0.2672|K|$ for the first-neighbor $\mathbb{Z}_2$-flux pair of Figs.~\ref{fig:Z-type}, \ref{fig:P-type} and $\Delta^{\prime} \approx 0.2372|K|$ for the second-neighbor $\mathbb{Z}_2$-flux pair of Fig.~\ref{fig:G-type} \cite{Kitaev2006}.  
The non-vanishing second-order perturbation terms obtained from Eq.~\eqref{EQ:perturb} should consist of two-, three-, four-spin interaction terms, up to the projection onto the zero-flux sector, obtained from the product of a pair of terms from $H_1$ that preserves all $\mathbb{Z}_2$ flux $W_p$. 
The resulting $H_{\rm eff}$ would be in the form of
\begin{widetext}
\begin{equation}
H_{\rm eff}
=
P_0 \left( 
K_1 \sum_{\langle jk \rangle_\gamma} \sigma_j^\gamma \sigma_k^\gamma
+
K_2 \sum_{\langle jkl \rangle_{\alpha\beta}} \sigma_j^\alpha \sigma_k^\gamma \sigma_l^\beta +
 \sum_{\langle jklm \rangle_{\alpha\gamma\beta}}  K_3^{\alpha\beta\gamma}\sigma_j^\alpha \sigma_k^\beta \sigma_l^\alpha \sigma_m^\beta
+
K_4 \sum_{\langle jklm \rangle_{\alpha\beta\alpha}} \sigma_j^\alpha \sigma_k^\gamma \sigma_l^\gamma \sigma_m^\alpha 
\right) P_0,
\label{EQ:perturbSpin}
\end{equation}
\end{widetext}
where $\langle jkl \cdots \rangle_{\alpha\gamma\cdots}$ denotes consecutive sites connected by the $\alpha$-bond between sites $j,k$, the $\gamma$-bond between sites $k,l$ {\it etc}, 
and the three Greek indices are all different from each other ($\alpha\ne\beta\ne\gamma$).
The effective spin interactions appearing in $H_{\rm eff}$ are essentially products of Kitaev bond interactions as they are the only terms that can survive under the projection operator $P_0$;
$$
\sigma_j^\alpha \sigma_k^\gamma \sigma_l^\beta \propto (\sigma_j^\alpha \sigma_k^\alpha)(\sigma_k^\beta \sigma_l^\beta) ,
$$
$$
\sigma_j^\alpha \sigma_k^\beta \sigma_l^\alpha \sigma_m^\beta 
 \propto (\sigma_j^\alpha \sigma_k^\alpha)(\sigma_k^\gamma \sigma_l^\gamma)(\sigma_l^\beta \sigma_m^\beta) ,
$$
$$
\sigma_j^\alpha \sigma_k^\gamma \sigma_l^\gamma \sigma_m^\alpha 
 \propto (\sigma_j^\alpha \sigma_k^\alpha)(\sigma_k^\beta \sigma_l^\beta)(\sigma_l^\alpha \sigma_m^\alpha) .
$$
See Fig.~\ref{fig:multi-spin-int} for a visualization of the effective spin interactions. 
We will later show that the spin interaction parameters come out to be 
\begin{equation}
K_1=K-\frac{2h^2}{3 \Delta}+\frac{2D_2^2}{\Delta'},
\end{equation}
\begin{equation}
K_2=\frac{4h\varGamma'}{\sqrt{3}\Delta},
\label{eq:K2}
\end{equation}
\begin{equation}
K_3^{\alpha\beta\gamma}=-\frac{2\left(\varGamma^\prime +\varepsilon_{\alpha\beta\gamma}D_1\right)^2}{\Delta},
\end{equation}
\begin{equation}
K_4=\frac{2(D_1^2-\varGamma'^2)}{\Delta},
\end{equation}
where $\varepsilon_{\alpha\beta\gamma}$ is the antisymmetric Levi-Civita tensor.

But $H_{\rm eff}$ can also be expressed as the hopping of Majorana fermions of the range varying from the nearest neighbor to the fourth neighbor \cite{Chari2021, Nasu2023, Takikawa2019, Fujimoto2020}. This implies that the third- and fourth-neighbor hopping is of the same order of magnitude as the second-neighbor hopping. Given that study of an analogous model based on graphene and the Haldane model have found topological phase transitions \cite{Sticlet2013, Fujimoto2020}, we shall now proceed to derive each non-vanishing second-order perturbation term.


\subsubsection{2- and 3-Spin Interactions}

We shall first discuss how to derive the first two terms of Eq.~\eqref{EQ:perturbSpin}. It would be relatively less complicated to follow in the sense that only 2 or 3 spins are involved. However, these terms do not represent the qualitatively more important part of our results; for the reasons discussed below, they do not drive the topological phase transitions.

We have stated through Eq.~\eqref{EQ:perturbSpin} that among the second-order perturbation terms, 
some are of the same form as $H_0$; they are hence of the lesser importance qualitatively. 
These terms can be attributed to two different sources. One comes from products of two Zeeman-type terms, or $h^2$-terms, 
that preserve the $\mathbb{Z}_2$-flux 
\cite{Kitaev2006}:
\begin{eqnarray}
        H_{\rm ZZ} &=& -\frac{1}{\Delta}P_0 \left[-h\sum_j\left(\frac{1}{\sqrt{3}}\sigma^x_j+\frac{1}{\sqrt{3}}\sigma^y_j+\frac{1}{\sqrt{3}}\sigma^z_j\right)\right]^2P_0\nonumber\\
        &=& -\frac{2h^2}{3\Delta}\!\sum_{\langle A,B \rangle_\gamma} P_0 
        \sigma_A^\gamma \sigma_B^\gamma P_0.
\label{EQ:ZZ}
\end{eqnarray}
The other comes from products of two $D_2$-type terms, 
for the derivation of which we can first consider only on the 
$x$-bond 
\begin{equation*}
    \begin{split}
        H_{D_2 D_2}^x = & -\frac{1}{\Delta^{\prime}}\sum_{\langle A,B \rangle_x}P_0 \left[D_2 \left(\sigma^y_A \sigma^z_B - \sigma^z_A \sigma^y_B\right)\right]^2P_0\\
        = & 
        \frac{D_2^2}{\Delta^{\prime}} \sum_{\langle A,B \rangle_x}
        P_0\left( \sigma_A^y \sigma_B^z \sigma_A^z \sigma_B^y 
        + \sigma_A^z \sigma_B^y \sigma_A^y \sigma_B^z \right)P_0 \\ 
        = & \frac{2 D_2^2 }{\Delta^{\prime}} \sum_{\langle A,B \rangle_x} P_0 \sigma_A^x \sigma_B^x P_0.
    \end{split}
\end{equation*}
By repeating for the $y,z$-bonds, we can obtain
\begin{equation}
        H_{D_2 D_2} = 
        \!\sum_{\alpha \in \{ x, y, z \}}\! H_{D_2 D_2}^{\alpha} = \frac{2 D_2^2 }{\Delta^{\prime}} \sum_{\langle A, B \rangle_\alpha} P_0 \sigma^\alpha_A \sigma^\alpha_B P_0,
\label{EQ:DD}
\end{equation}
which completes our derivation of 
$K_1=K-\frac{2h^2}{3\Delta}+\frac{2D_2^2}{\Delta^\prime}$ in Eq.~\eqref{EQ:perturbSpin}

The next term in Eq.~\eqref{EQ:perturbSpin} is the 3-spin interaction whose parameter $K_2$ as we shall show is 
nonzero only for $h \neq 0$. It is obtained at the second order of perturbation in our model 
as products of a single Zeeman-type term with a single $D_1/\varGamma^{\prime}$-type term that preserve the $\mathbb{Z}_2$-flux. 
As a concrete example, we derive the 3-spin interactions in the following two cases.
\begin{widetext}
\begin{equation}
\includegraphics[height=2cm,valign=c]{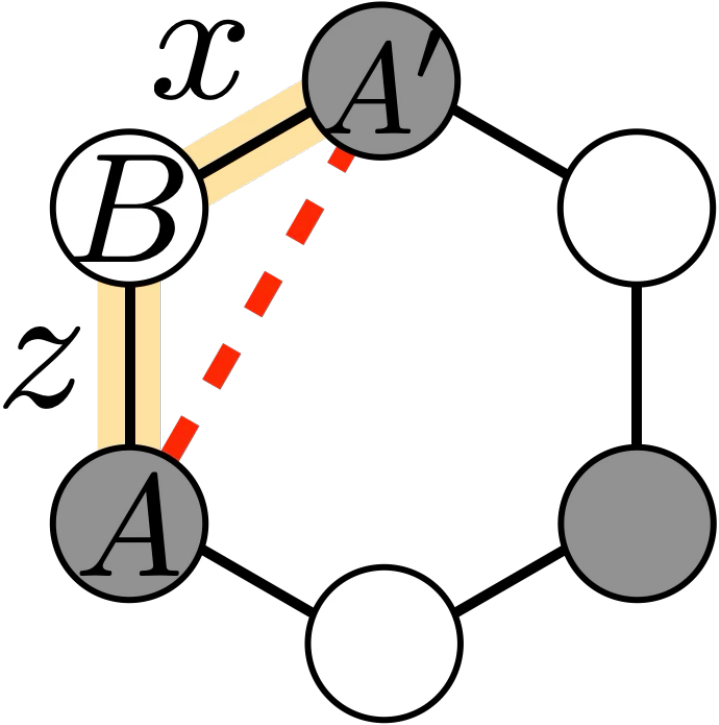}
~~~~~\langle ABA'\rangle_{zx}:~~~
\begin{array}{l}
\frac{2}{\Delta} P_0
\left[\frac{h}{\sqrt{3}} \sigma^z_{A} \right]
\left[(\varGamma^\prime+D_1) \sigma^y_B \sigma^x_{A'} \right] P_0
+
\frac{2}{\Delta} P_0 
\left[(\varGamma^\prime-D_1) \sigma^z_{A} \sigma^y_B \right]
\left[\frac{h}{\sqrt{3}}\sigma^x_{A'} \right] P_0
\\
= 
\frac{4h\varGamma^\prime }{\sqrt{3}\Delta} P_0 \sigma_{A}^z \sigma_{B}^y \sigma_{A^{\prime}}^x  P_0.
\end{array}
\label{EQ:SecA}
\end{equation}
\begin{equation}
\includegraphics[height=2cm,valign=c]{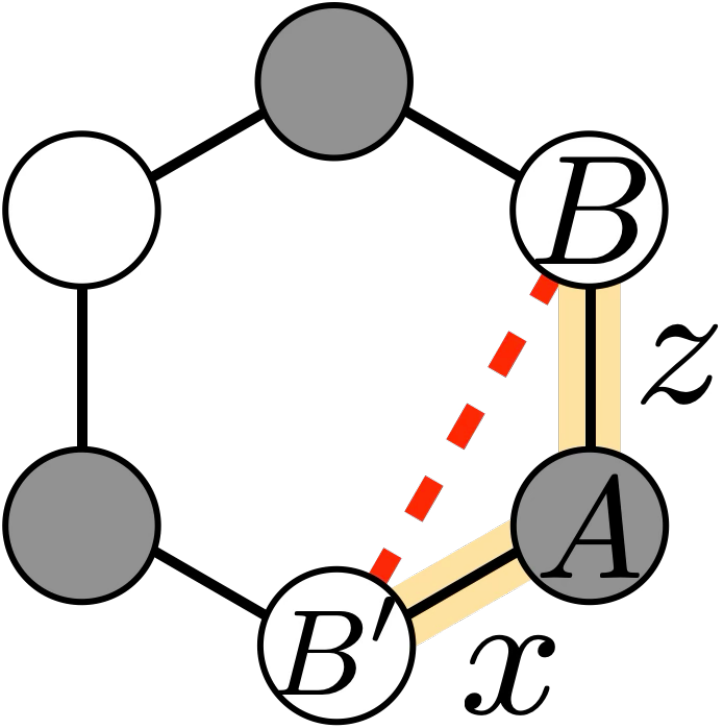}
~~~~~\langle BAB'\rangle_{zx}:~~~
\begin{array}{l}
\frac{2}{\Delta} P_0 \left[(\varGamma^\prime+D_1)\sigma^z_{B} \sigma^y_A \right] \left[\frac{h}{\sqrt{3}}\sigma^x_{B'} \right] P_0
+
\frac{2}{\Delta} P_0 
\left[\frac{h}{\sqrt{3}}\sigma^z_{B} \right] 
\left[ (\varGamma^\prime-D_1) \sigma^y_A \sigma^x_{B'} \right] P_0
\\
= 
\frac{4h\varGamma^\prime }{\sqrt{3}\Delta} P_0 \sigma_B^z \sigma_A^y \sigma_{B'}^x  P_0.
\end{array}
\label{EQ:SecB}
\end{equation}
\end{widetext}
By applying lattice translations and C$_3$ rotations to the two cases, we can generate all the 3-spin interactions and eventually obtain the whole 3-spin term:
\begin{equation}
H_{\rm eff}^{(2)}  
=\frac{4h\varGamma^\prime }{\sqrt{3}\Delta}\sum_{\langle jkl \rangle_{\alpha\beta}}P_0\sigma^\alpha_j \sigma^\gamma_k \sigma^\beta_l P_0,
\label{EQ:ZG}
\end{equation}
completing our derivation of 
$K_2=\frac{4h\varGamma^\prime}{\sqrt{3}\Delta}$ 
in Eq.~\eqref{EQ:perturbSpin}. 
With the regards the magnitude of the 3-spin interaction, we will focus on the small Zeeman field regime, {\it i.e.} $|h| \ll |D_1|, |\varGamma^\prime|$, and hence it will not be as large as that of the 4-spin interaction which we will derive next. 
Nonetheless, we shall later show how this 3-spin interaction can not be ignored 
for any 
$h\varGamma^\prime \neq 0$.

\subsubsection{4-Spin Interactions}

We can now show that the principal effect of the second-order perturbation arises from the induced 4-spin interaction terms. 
This is because the terms with the parameters quadratic in $D_1$ and $\varGamma^\prime$ 
contribute 
only to 
the 4-spin interaction parameters $K_3^{\alpha\beta\gamma}$ and $K_4$ in Eq.~\eqref{EQ:perturbSpin}. 
The 4-spin term with the parameter $K_3^{\alpha\beta\gamma}$ can be obtained by second-order perturbation on the following types of three consecutive bonds:
\begin{widetext}
\begin{equation}
\includegraphics[height=2cm,valign=c]{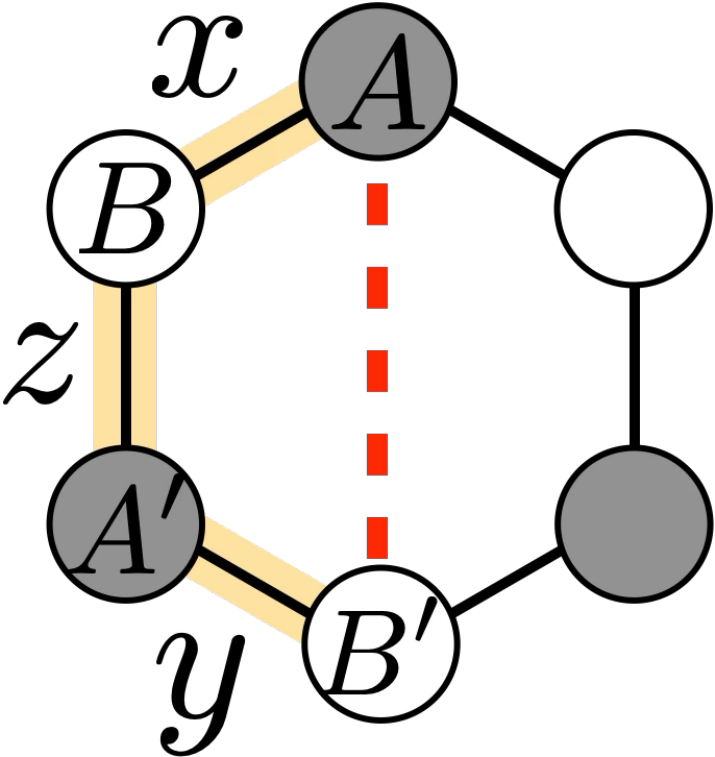}
~~~~~\langle ABA'B'\rangle_{xzy}:~~~
\begin{array}{l} 
-\frac{2}{\Delta} 
P_0\left[ 
        \left( \varGamma^{\prime} + D_1 \right) 
        \sigma_{A}^x \sigma_{B}^y 
        \right]
        \left[\left( \varGamma^{\prime} + D_1 \right)
        \sigma_{A^{\prime}}^x \sigma_{B^{\prime}}^y 
\right] P_0
\\
=
- \frac{2\left( \varGamma^{\prime} + D_1 \right)^2}{\Delta} P_0 \sigma_{A}^x \sigma_{B}^y \sigma_{A^{\prime}}^x \sigma_{B^{\prime}}^y P_0 ,
\end{array}
\label{EQ:D1p}
\end{equation}
\begin{equation}
\includegraphics[height=2cm,valign=c]{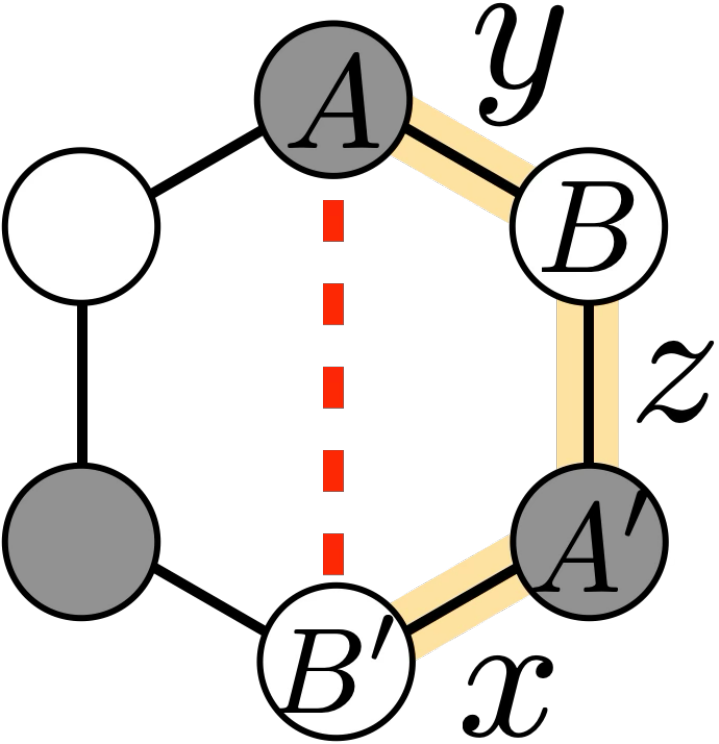}
~~~~~\langle ABA'B'\rangle_{yzx}:~~~
\begin{array}{l}
-\frac{2}{\Delta} 
P_0\left[ 
\left( \varGamma^{\prime} - D_1 \right) 
        \sigma_{A}^y \sigma_{B}^x 
        \right]
        \left[\left( \varGamma^{\prime} - D_1 \right)
        \sigma_{A^{\prime}}^y \sigma_{B^{\prime}}^x 
\right] P_0
\\
=
- \frac{2\left( \varGamma^{\prime} - D_1 \right)^2}{\Delta} P_0 \sigma_{A}^y \sigma_{B}^x \sigma_{A^{\prime}}^y \sigma_{B^{\prime}}^x P_0.
\end{array}
\label{EQ:D1m}
\end{equation}
\end{widetext}
Note that the two different paths connecting the same end points ($A$ and $B'$) lead to different results, which are manifestations of the C$_2$ rotation symmetry 
as shown in Fig.~\ref{fig:lattice-symm} 
broken by the DM interactions as stated in Tab.~\ref{tab:symm}. 
Collecting all the 4-spin interactions related by lattice translations and C$_3$ rotations, we obtain the whole 4-spin term of this type 
\begin{equation}
H_{\rm eff}^{(3)}
= -\sum_{\langle jklm \rangle_{\alpha\gamma\beta}}  \frac{2\left(\varGamma^\prime  + \varepsilon_{\alpha\beta\gamma} D_1\right)^2}{\Delta}P_0\sigma^\alpha_j \sigma^\beta_k \sigma^\alpha_l \sigma^\beta_m P_0 ,
\label{EQ:GG3}
\end{equation}
where we again take the convention that $j$ and $l$ sites belong to the $A$ sublattice but $k$ and $m$ sites belong to the $B$ sublattice; this completes our derivation of $K_3^{\alpha\beta\gamma}=-\frac{2(\varGamma^\prime+\varepsilon_{\alpha\beta\gamma}D_1)^2}{\Delta}$ in Eq.~\eqref{EQ:perturbSpin}. 
The 4-spin term with the parameter $K_4$ can also be 
computed in the similar 
fashion. 
Here we consider the following three consecutive bonds: 
\begin{widetext}
\begin{equation}
\includegraphics[height=2cm,valign=c]{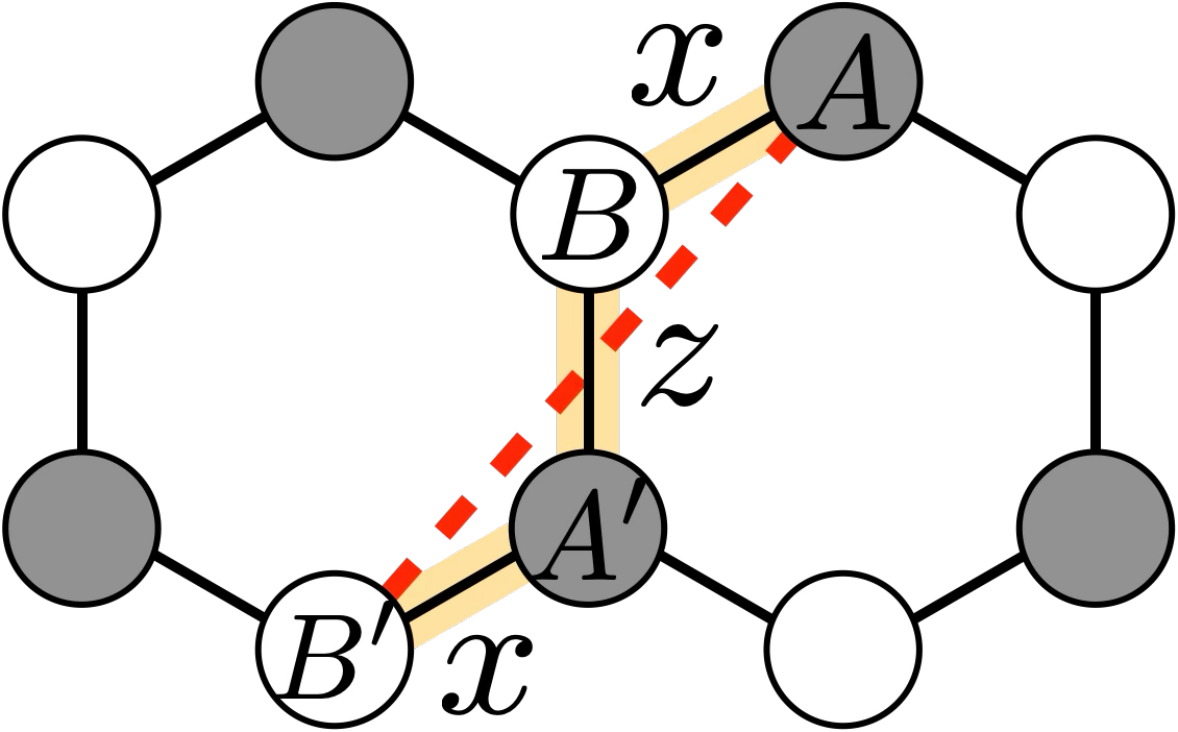}
~~~~~\langle ABA'B' \rangle_{xzx}:~~~
\begin{array}{l}
-\frac{2}{\Delta}
P_0 \left[ 
(\varGamma'+D_1) \sigma_{A}^x \sigma_{B}^y 
\right]
\left[(\varGamma'-D_1) \sigma_{A'}^y \sigma_{B'}^x 
\right]P_0
\\
= 
-\frac{2(\varGamma'^2-D_1^2)}{\Delta}
P_0
\sigma_{A}^x \sigma_{B}^y 
\sigma_{A'}^y \sigma_{B'}^x 
P_0.
\end{array}
\end{equation}
\begin{equation}
\includegraphics[height=2cm,valign=c]{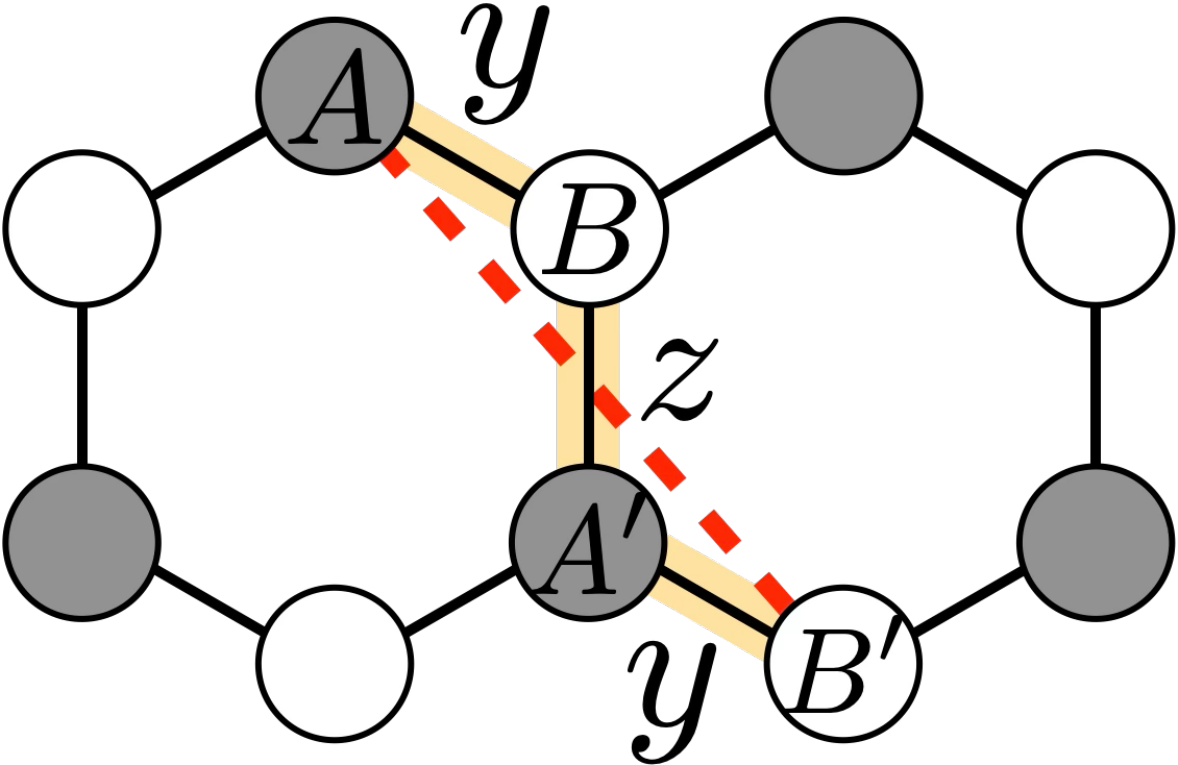}
~~~~~\langle ABA'B' \rangle_{yzy}:~~~
\begin{array}{l}
-\frac{2}{\Delta}
P_0 \left[ 
(\varGamma'-D_1) \sigma_{A}^y \sigma_{B}^x 
\right]
\left[(\varGamma'+D_1) \sigma_{A'}^x \sigma_{B'}^y 
\right]P_0
\\
= 
-\frac{2(\varGamma'^2-D_1^2)}{\Delta}
P_0
\sigma_{A}^y \sigma_{B}^x 
\sigma_{A'}^x \sigma_{B'}^y 
P_0.
\end{array}
\end{equation}
\end{widetext}
Interestingly, these two cases give identical results since they are related by the mirror symmetry that is still preserved by the DM interactions. Because of this property, the whole 4-spin term this type 
does not possess any particular bond dependence: 
\begin{equation}
H_{\rm eff}^{(4)} 
= 
\frac{2\left(D_1^2-\varGamma'^2\right)}{\Delta}\sum_{\langle jklm \rangle_{\alpha\beta\alpha}} P_0 \sigma^\alpha_j \sigma^\gamma_k \sigma^\gamma_l \sigma^\alpha_m P_0,
\label{EQ:GG4}
\end{equation}
where we take $j$ and $l$ sites from the same sublattice but $k$ and $m$ sites from the other sublattice; this completes our derivation of $K_4=\frac{2(D_1^2-{\varGamma^\prime}^2)}{\Delta}$ in Eq.~\eqref{EQ:perturbSpin}.

\subsection{Topology analysis from Majorana 
Hamiltonian}

As our main purpose is to analyze the topological characteristics of $H_{\rm eff}$ [Eq.~\eqref{EQ:perturbSpin}], which can be carried out simplest from the Majorana fermion representation in the momentum space, the logical next step is to obtain its Majorana fermion representation 
in the real space, where the 3- and 4-spin interactions can be expressed as longer-ranged Majorana hoppings. We note that in the Majorana representation the zero-flux projection operator $P_0$  can be taken care of through the gauge fixing
$$
i b^\gamma_A b^\gamma_B = +1
$$
when A, B are nearest-neighbor sites belonging to the A- and B-sublattice, respectively; this sets $W_p = +1$ for all plaquettes from the Eq.~\eqref{EQ:fluxMajorana} Majorana representation $W_p = \prod_{\langle A,B \rangle_\gamma} \left(ib_A^\gamma b_B^\gamma\right)$. Applying this gauge fixing and the Eq.~\eqref{EQ:spinMajorana} Majorana representation of spin-1/2 operators, $\sigma^\gamma_j = ib^\gamma_j c_j$ after re-writing all the spin interaction of $H_{\rm eff}$ as products of the Kitaev spin interaction, we obtain
\begin{widetext}
\begin{equation}
H_{\rm eff} = -iK_1 \sum_{\langle A,B\rangle_1} c_A c_B -i K_2 \left(\sum_{\langle A, A^\prime\rangle_2} c_A c_{A^\prime}+\sum_{\langle B, B^\prime\rangle_2} c_B c_{B^\prime}\right)+i\left(K^{xyz}_3+K_3^{yxz}\right)\sum_{\langle A,B\rangle_3} c_A c_B -iK_4 \sum_{\langle A, B \rangle_4} c_A c_B,
\label{EQ:realSpaceMajorana}
\end{equation}
\end{widetext}
where $\langle A, B\rangle_n$ denotes the $n$-th neighbor A, B belonging to the A-, B-sublattice respectively and $\langle A (B), A^\prime (B^\prime)\rangle_2$ denotes the 2nd neighbor pairs both belonging to the A(B)-sublattice in the counter-clockwise direction.

\begin{figure}[b]
    \centering
    \includegraphics[width=\linewidth]{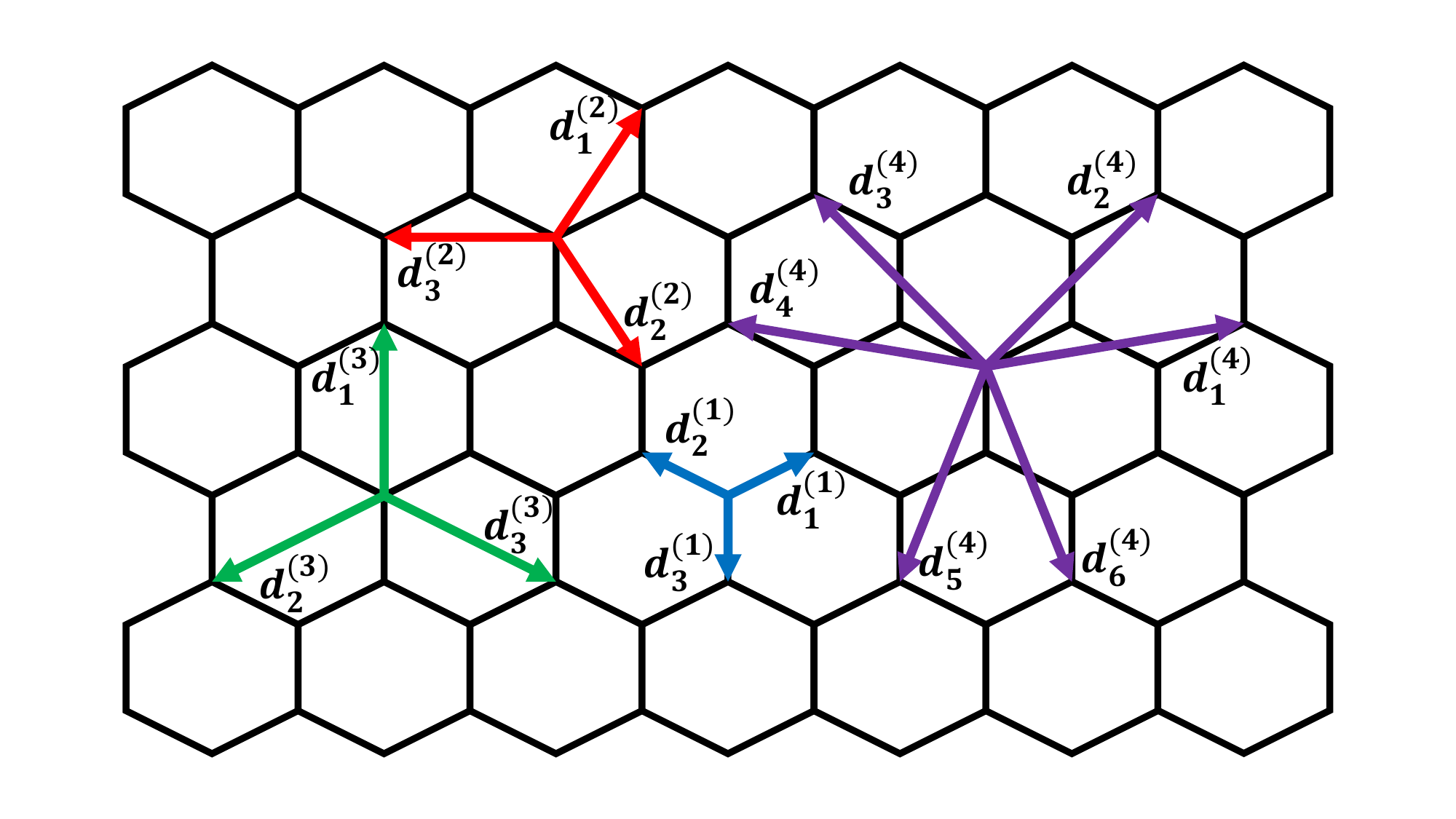}
    \caption{The lattice vectors are defined as follows: ${\bf d}_m^{(1)}$, represented by a blue arrow, is the nearest-neighbor lattice vector. ${\bf d}_m^{(2)}$, represented by a red arrow, is the second-neighbor lattice vector. ${\bf d}_m^{(3)}$, represented by a green arrow, is the third-neighbor lattice vector. ${\bf d}_m^{(4)}$, represented by a purple arrow, is the fourth-neighbor lattice vector.}
    \label{fig:Lattice_vector}
\end{figure}

Next we transform the above real-space Hamiltonian of Eq.~\eqref{EQ:realSpaceMajorana} into the momentum-space Hamiltonian. 
We first note that the first term of Eq.~\eqref{EQ:realSpaceMajorana} is a Majorana fermion analogue of the nearest-neighbor hopping only graphene model. 
Hence, the Majorana cones (MCs) are 
present at the ${\rm K}$ and ${\rm K'}$ points in the first Brillouin zone, 
analogous to the Dirac cones of graphene \cite{Kitaev2006, Motome2020}. We have seen that 
the electric field induced 
DM interaction 
and the trigonal distortion induced $\varGamma^\prime$ interaction 
effectively give rise to additional distant hopping terms through the perturbative treatment of off-diagonal spin interactions \cite{Chari2021, Takikawa2019}. This system therefore resembles graphene with distant-neighbor hopping \cite{Sticlet2013}. To show this, we first define the Fourier transformation of the Majorana fermion as
\begin{equation}
    c_{{\bf r},\rho} = \frac{1}{\sqrt{N}}\sum_{\bf k} e^{i {\bf k} \cdot {\bf r}} c_{{\bf k}, \rho} 
\end{equation}
where ${\bf r}$ is the position vector of lattice site, $\rho (\in { A, B })$ denotes the sublattice, and ${\bf k}$ is crystal momentum. The lattice vectors  $\{{\bf d}_m^{(n)}\}$ connecting $n$-th neighboring sites associated with the $K_n(n=1,2,3,4)$ interaction terms are depicted in Fig.~\ref{fig:Lattice_vector}.
By applying the Fourier transformation, the entire Hamiltonian can be expressed as
\begin{equation}
    H_{\rm eff} = \frac{1}{2}\sum_{{\bf k}}
    \begin{pmatrix}
        c_{-{\bf k}, A} & c_{-{\bf k}, B}
    \end{pmatrix}
    \begin{pmatrix}
        M ({\bf k}) & f({\bf k}) \\
        f^* ({\bf k}) & -M({\bf k})
    \end{pmatrix}
    \begin{pmatrix}
        c_{{\bf k}, A} \\
        c_{{\bf k}, B}
    \end{pmatrix}
    \label{EQ:kEff}
\end{equation}
where $f({\bf k}) \equiv f_1 ({\bf k}) + f_3 ({\bf k}) + f_4 ({\bf k})$ with
\begin{equation}
    f_1 ({\bf k}) = -i K_1 \sum_{m=1}^3 e^{-i {\bf k} \cdot {{\bf d}_m^{(1)}}},
\end{equation}
\begin{equation}
    f_3 ({\bf k}) = i \left(K^{xyz}_3+K_3^{yxz}\right) \sum_{m=1}^3 e^{-i {\bf k} \cdot {{\bf d}_m^{(3)}}},
    \label{EQ:hop3}
\end{equation}
\begin{equation}
    f_4 ({\bf k}) = -i K_4 \sum_{m=1}^6 e^{-i {\bf k} \cdot {{\bf d}_m^{(4)}}},
    \label{EQ:hop4}
\end{equation}
and
\begin{equation}
    M({\bf k}) = - K_2 \sum_{m=1}^3 \sin \left[{\bf k} \cdot {{\bf d}_m^{(2)}}\right] .
    \label{EQ:MFS}
\end{equation}
Note that the Majorana mass term $M({\bf k})$ becomes nonzero only in presence of a magnetic field~[Eq.~\eqref{eq:K2}].
The Majorana Hamiltonian $H_{\rm eff}$ [Eq.~\eqref{EQ:kEff}] 
gives the energy dispersion of Majorana fermion:
\begin{equation}
\epsilon({\bf k}) = 
\pm \sqrt{|f({\bf k})|^2+M^2 ({\bf k})}.
\end{equation}
We shall first consider the $h=0$ case 
and then move on to the $h \neq 0$ case.

\subsubsection{Additional Majorana cones for $h=0$}

The $h=0$ case is closely analogous to the graphene with the long-range hopping, 
which implies, for strong enough third- and the fourth-neighbor hopping \cite{Sticlet2012, Sticlet2013, Fujimoto2020}, the possibility of additional Majorana cones (AMCs). 
For this case, the effective 
Hamiltonian of Eq.~\eqref{EQ:kEff} can be represented 
by the matrix
\begin{equation}
    \begin{pmatrix}
        0 & f({\bf k}) \\
        f^* ({\bf k}) & 0
    \end{pmatrix} = \boldsymbol{\mathcal{F}} ({\bf k}) \cdot \boldsymbol {\sigma} ,
    \label{EQ:MajoranaHeff}
\end{equation}
where $\boldsymbol{\mathcal{F}} ({\bf k}) = [{\rm Re}f({\bf k}), -{\rm Im}f({\bf k}), 0]$ and $\boldsymbol{\sigma}=[\sigma_x,\sigma_y,\sigma_z]$ ($\sigma_{x,y,z}$ are the Pauli spin matrices). 
AMCs can be identified by solving the equation of $f({\bf k})=0$; more explicitly,
\begin{eqnarray}
\sum_{m=1}^3 e^{-i {\bf k} \cdot {\bf d}_m^{(1)}} + 2\zeta \sum_{m=1}^3 e^{-i {\bf k} \cdot {\bf d}_m^{(3)}} + \eta \sum_{m=1}^6 e^{-i {\bf k} \cdot {\bf d}_m^{(4)}} = 0,~~~~~~~
\label{EQ:MCs}
\end{eqnarray}
where the two parameters,
\begin{eqnarray}
\zeta &=& -\frac{K^{xyz}_3+K_3^{yxz}}{2K_1} = \frac{2\left(D_1^2 + {\varGamma^\prime}^2\right)/\Delta}{K+ 2D_2^2/\Delta'},
\label{EQ:zeta}
\\
\eta &=& \frac{K_4}{K_1} = \frac{2\left(D_1^2 - {\varGamma^\prime}^2\right)/\Delta}{K+ 2D_2^2/\Delta'},
\label{EQ:eta}
\end{eqnarray}
determine the number of AMCs and their topological structures.
Note 
our perturbation theory 
gives us $\zeta/\eta = (D_1^2 + {\varGamma^\prime}^2)/(D_1^2 - {\varGamma^\prime}^2)$, and 
hence 
$|\zeta| \geq |\eta|$.

\begin{figure}[tb]
    \centering
    \includegraphics[width=0.8\linewidth]{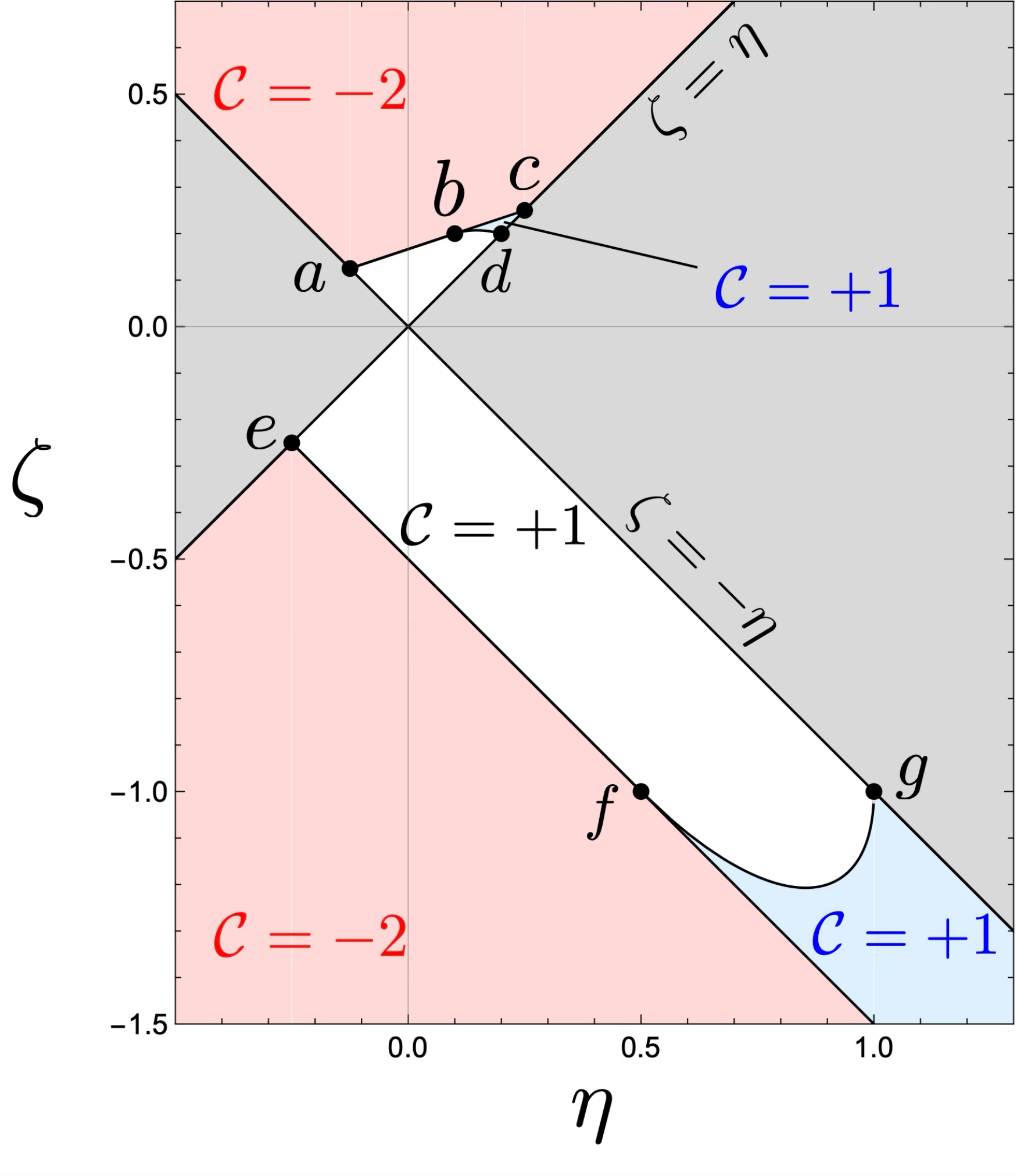}
    \caption{
    Phase diagram of the Majorana Hamiltonian [Eq.~(\ref{EQ:realSpaceMajorana})] in the parameter space of $(\eta, \zeta)$.
    The colored regions represent the number of zero-field Majorana cones or the number of solutions of Eq.~(\ref{eq:AMC_on_kxline}):
    (i) 7 MCs near the K valley or $\{x_{\rm K},x_+,x_-\}$ (blue), (ii) 4 MCs near the K valley or $\{x_{\rm K},x_+\}$/$\{x_{\rm K},x_-\}$ (red), and (iii) a single MC near the K valley or $\{x_{\rm K}\}$ (white).
    The gray region is excluded from our consideration due to the condition, $|\zeta|\ge|\eta|$, deduced from our perturbation theory.
    The applied magnetic field gaps out the zero-field MCs in each case, resulting in the nonzero Chern number: (i) $\mathcal{C}=+1$ (blue), (ii) $\mathcal{C}=-2$ (red), and (iii) $\mathcal{C}=+1$ (white).
    The marked points are:
    $a=(-1/8,1/8)$, $b=(1/10,1/5)$, $c=(1/4,1/4)$, $d=(1/5,1/5)$, $e=(-1/4,-1/4)$, $f=(1/2,-1)$, and $g=(1,-1)$.
    The explicit expressions of the boundary lines are provided in Eqs.~(\ref{EQ:realAMC}),~(\ref{EQ:pRange}),~and~(\ref{EQ:mRange}) of Appendix~\ref{APP:pmExplicit}.
    Along the boundary lines, the zero-field system exhibits nontrivial Majorana band topology such as quadratic band touching and line node (discussed in Figs.~\ref{fig:zeta=-eta},~\ref{fig:eta=0},~\ref{fig:zetaNOT=eta},~and~\ref{fig:zeta=eta}).
    The chirality structures of the zero-field MCs are illustrated in Fig.~\ref{fig:AMCs}.
    }
    \label{fig:chirality}
\end{figure}

\begin{figure*}
     \centering
      \begin{subfigure}[b]{0.23\linewidth}
         \centering
         \includegraphics[width=\linewidth]{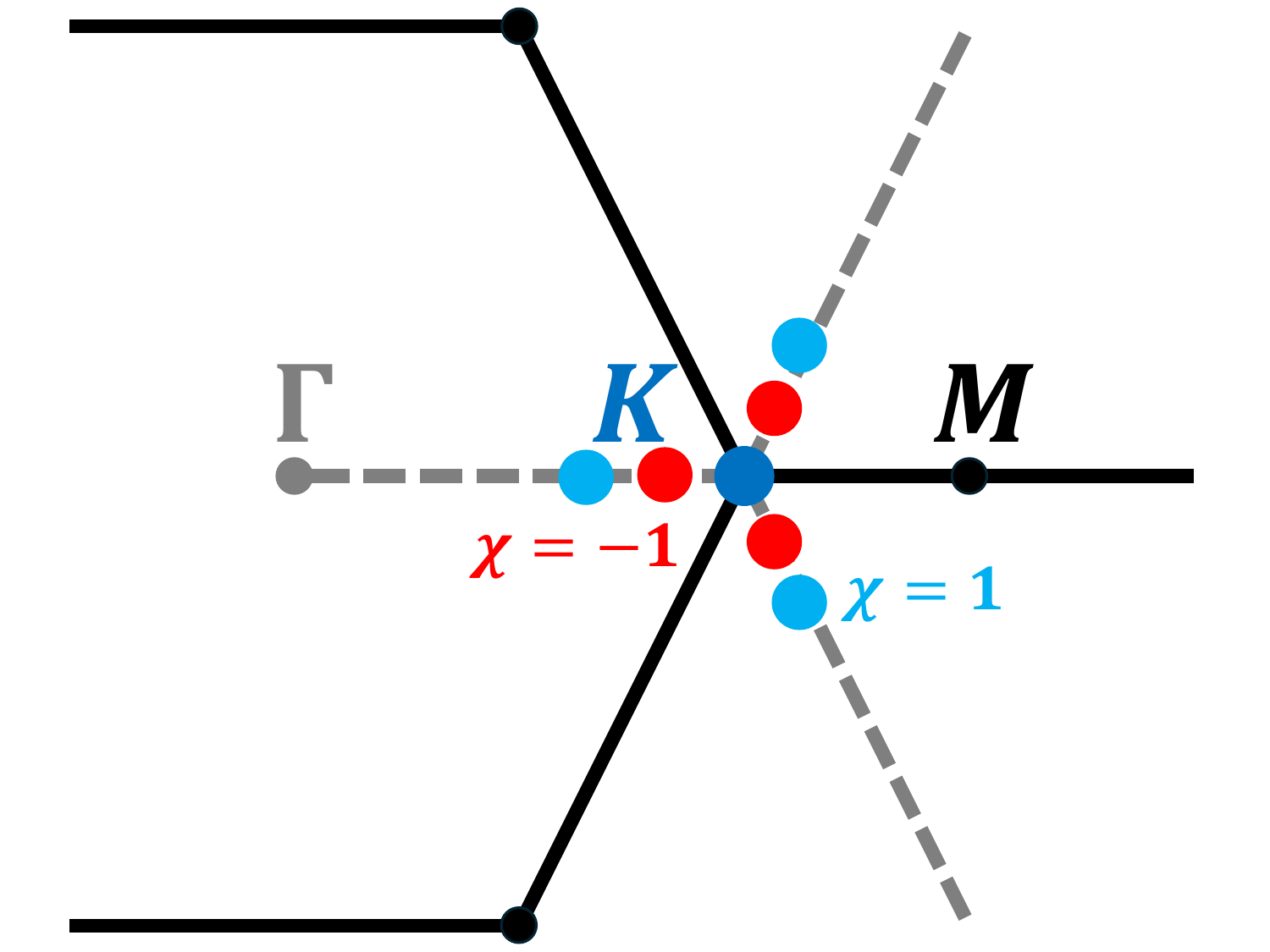}
         \caption{}
         \label{fig:AMC1}
     \end{subfigure}
     \hfill
    \begin{subfigure}[b]{0.23\linewidth}
         \centering
         \includegraphics[width=\linewidth]{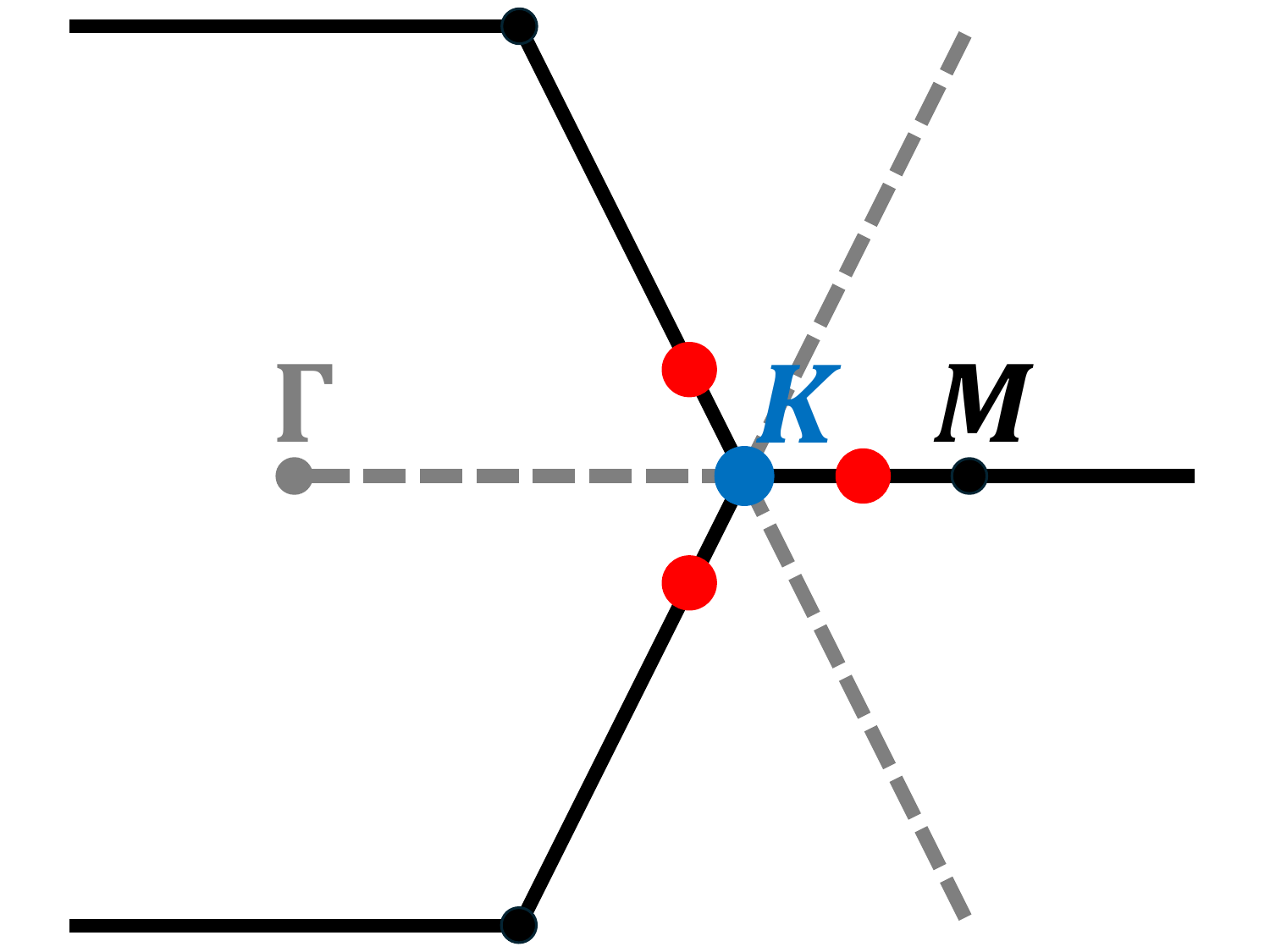}
         \caption{}
         \label{fig:AMC2}
    \end{subfigure}
    \hfill
     \begin{subfigure}[b]{0.23\linewidth}
         \centering
         \includegraphics[width=\linewidth]{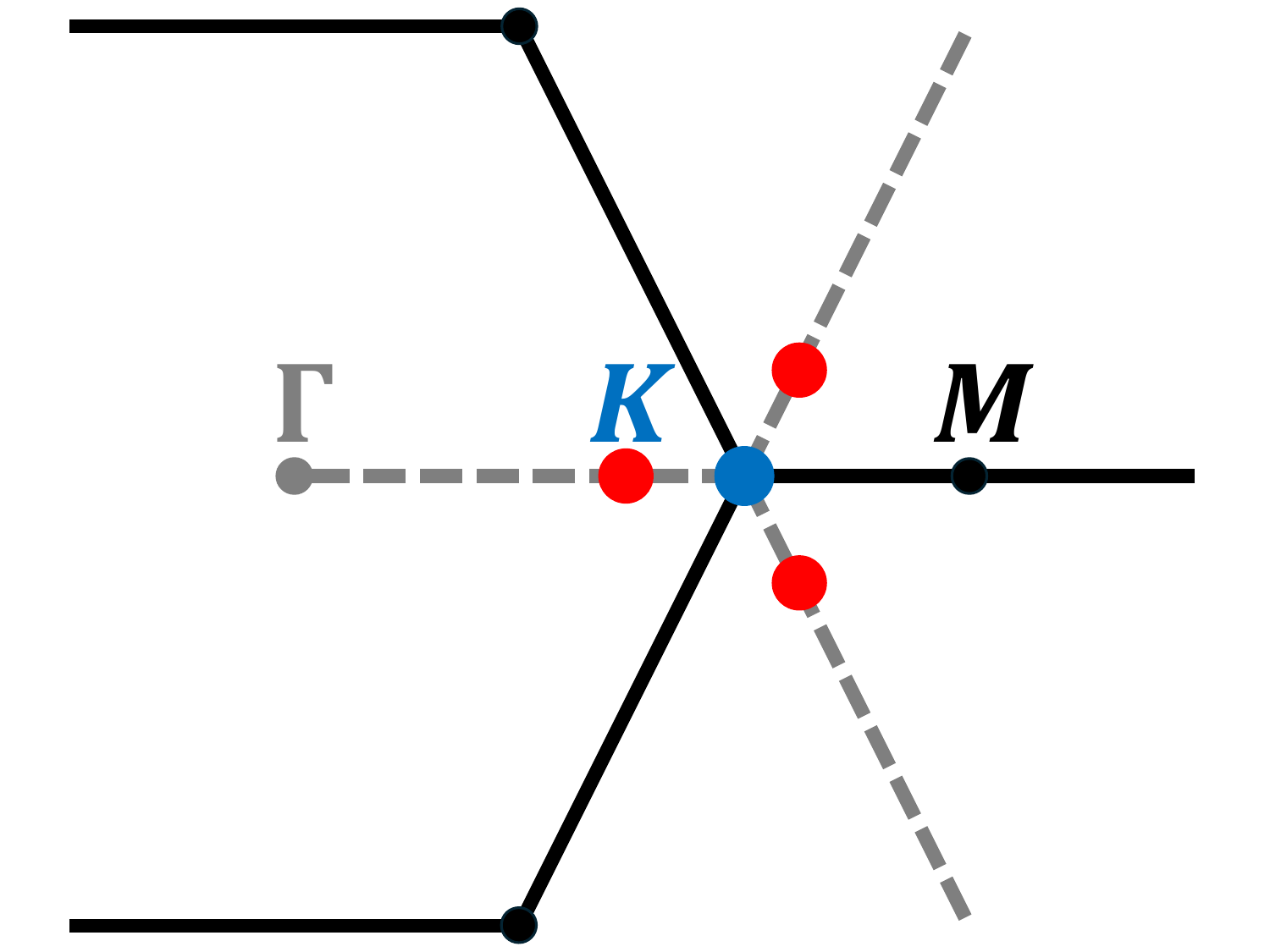}
         \caption{}
         \label{fig:AMC3}
     \end{subfigure}
     \hfill
     \begin{subfigure}[b]{0.23\linewidth}
         \centering
         \includegraphics[width=\linewidth]{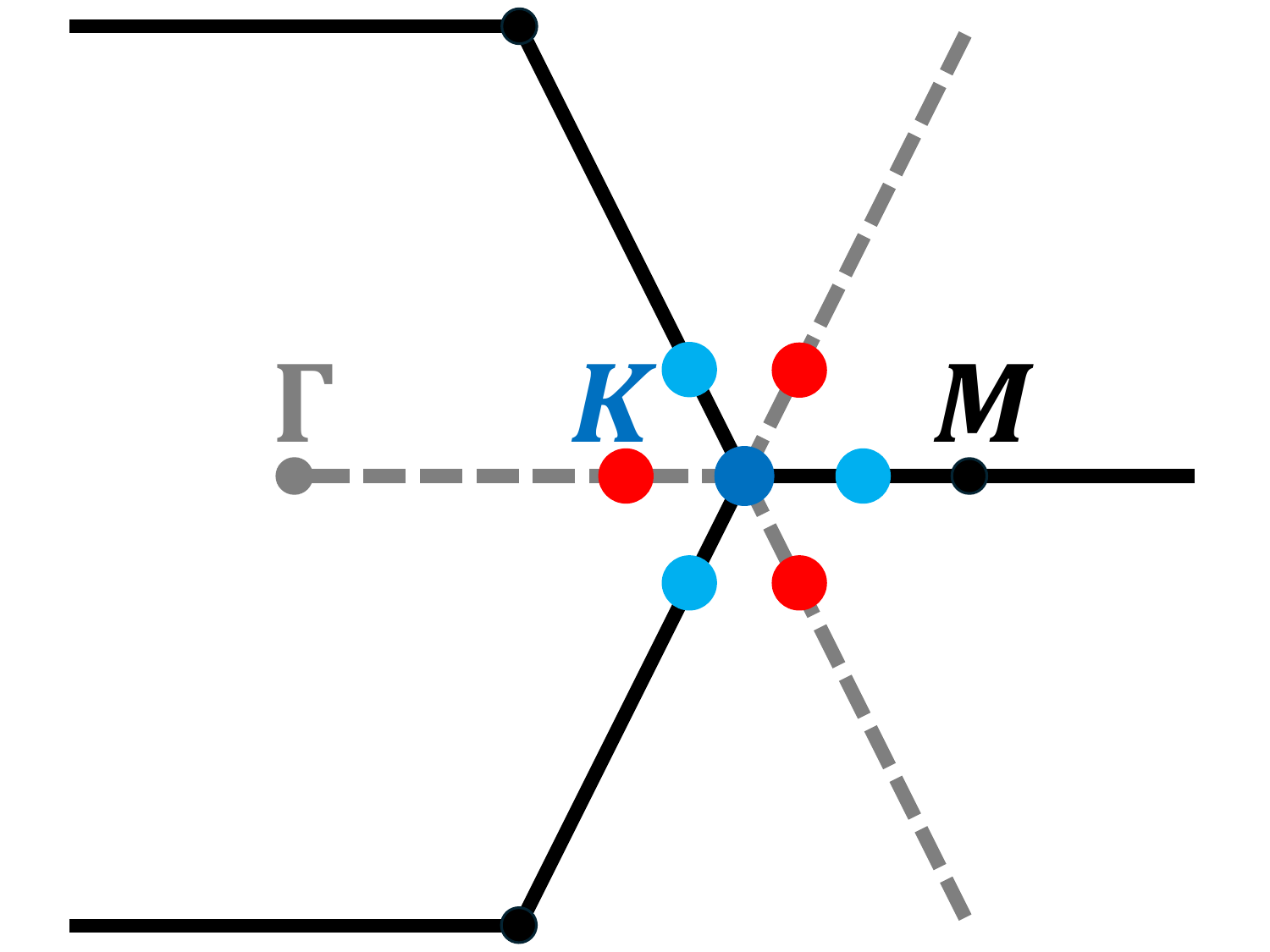}
         \caption{}
         \label{fig:AMC4}
     \end{subfigure}
    \caption{
Some characteristic AMC distribution near the ${\rm K}$ point. The dark blue dot represents the MC at ${\rm K}$-point with the positive chirality. 
Light blue(red) dots represent AMCs with 
the positive(negative) chirality. 
We 
take $M({\bf k})>0$ in the ${\rm K}$ valley. 
(a) The case of $\zeta < -1$ in the $D_1=0$ limit has the Chern number $C=+1$. (b) The case of $\frac{1}{8}<\zeta < \frac{1}{3}$ in the $D_1=0$ limit has the Chern number $C=-2$.  
    (c)  The case of $\zeta > \frac{1}{3}$ in the $D_1=0$ limit and $\zeta < -\frac{1}{4}-\frac{\delta}{8}$ or $\zeta>\frac{1}{4}-\frac{\delta}{8}$ in the $\varGamma^\prime \ll |D_1|$ limit has the Chern number $C=-2$. (d) The case of $\frac{1}{5}+\frac{\delta}{25}<\zeta < \frac{1}{4}-\frac{\delta}{8}$ in the $\varGamma^\prime \ll |D_1|$ limit has the Chern number $C=+1$.}
    \label{fig:AMCs}
\end{figure*}

Any AMCs that arise are located on the 3 high-symmetry lines of the 1st BZ joining $\Gamma$, ${\rm K}$, ${\rm M}$, and ${\rm K'}$. 
This can be shown analytically for $\zeta \to \pm\infty$ or $\eta \to \pm \infty$ cases \cite{Sticlet2013}. The latter is particularly simple, as one can read off $\sum_{m=1}^3 \exp\left[-i{\bf k}\cdot{\bf d}_m^{(3)}\right]=\sum_{m=1}^3 \exp\left[i2{\bf k}\cdot{\bf d}_m^{(1)}\right]$ from Fig.~\ref{fig:Lattice_vector}; this implies that 
the midpoints of the $\Gamma{\rm K}$ and $\Gamma{\rm K'}$ lines are additional solutions for $\sum_{m=1}^3 \exp\left[-i{\bf k}\cdot{\bf d}_m^{(3)}\right]=0$.  Fig.~\ref{fig:AMCs} shows some characteristic examples of how the AMCs thus 
arises around ${\rm K}$. Given that $k_y=0$ for one such high-symmetry line, it would be sufficient to solve Eq.~\eqref{EQ:MCs} just for ${\bf k} = k_x{\bf \hat{x}}$, which gives us
\begin{equation}
(2x+1)[1+2\zeta(2x-1)+2\eta(2x^2-1)]=0,
\label{eq:AMC_on_kxline}
\end{equation}
where $x\equiv \cos \frac{\sqrt{3}}{2}k_x$.
An obvious solution of the above equation is
\begin{equation}
x_{\rm K}=-1/2,
\end{equation}
corresponding to the original MC at the ${\rm K}$ point.
There can be two more solutions,
\begin{equation}
x_\pm = \frac{-\zeta \pm \sqrt{\zeta^2-\eta(1-2\zeta-2\eta)}}{2\eta},
\label{EQ:xPM}
\end{equation}
provided that the two conditions, $x_{\pm}\in \mathbb{R}$ and $| x_\pm | \leq 1$, are satisfied. The explicit constraints on $\zeta$ and $\eta$ due to the two conditions are provided in Eqs.~(\ref{EQ:realAMC}),~(\ref{EQ:pRange}),~and~(\ref{EQ:mRange}) of Appendix~\ref{APP:pmExplicit}. 
Depending on the parameter values of $\zeta$ and $\eta$, the solution of Eq.~\eqref{eq:AMC_on_kxline} corresponds to one of the three cases:
(i) $\{x_{\rm K},x_{+},x_{-}\}$,
(ii) $\{x_{\rm K},x_{+}\}$ or $\{x_{\rm K},x_{-}\}$,
and (iii) $\{x_{\rm K}\}$ (Fig.~\ref{fig:chirality}).

Although we have focused on AMCs along the $\Gamma{\rm K}$ line, there exist two more lines 
related to the $\Gamma {\rm K}$ line by threefold rotation symmetry about the ${\rm K}$ point. Hence, $x_+$ (and also $x_-$) actually gives a trio of AMCs related by the threefold rotation (see Fig.~\ref{fig:AMCs}). It must be remembered that the ${\rm K'}$ point should also have the same structure of AMCs due to time reversal symmetry. 


Each MC should not only have zero energy but also possess a well-defined chirality, which means the winding number of the vector $\boldsymbol{\mathcal{F}} ({\bf k})$ around the zero-energy point~\cite{Sticlet2013,SSZhang2020}.
The chirality of each MC is given by
\begin{equation}
    \chi = 
    \sgn \left[ \frac{1}{2\pi}\hat{\bf z}\cdot\int_{\rm MC} d^2 {\bf k}  \left\{\frac{\partial \hat{\boldsymbol{\mathcal F}} ({\bf k})}{\partial k_x} \times \frac{\partial \hat{\boldsymbol{\mathcal F}}({\bf k})}{\partial k_y}\right\} \right],
    \label{EQ:chiral1}
\end{equation}
where the 2D integration is performed over only single MC.
%
It is straightforward to obtain the latter from the first derivative of $\boldsymbol{\mathcal{F}} ({\bf k})$, which is required to be non-vanishing at an MC. Calculating this first derivative is further simplified for $k_y=0$ as that will give us $\partial_{k_x} \mathcal{F}_x = \partial_{k_y} \mathcal{F}_y = 0$, and Eq.~\eqref{EQ:chiral1} simplifies to
\begin{equation}
\chi = -{\rm sgn}[(\partial_{k_x} \mathcal{F}_y)(\partial_{k_y} \mathcal{F}_x)].
\end{equation}

From the chirality calculations, we obtain 
the following features of the topological structure of AMCs: 

\begin{itemize}
\item In the ${\rm K}$-valley, the chiralities of the MCs are determined by
\begin{eqnarray}
\chi (x_{\rm K}) &=& +1,
\\
\chi (x_\pm) &=& \mp {\rm sgn}(\zeta - \eta) = \mp {\rm sgn}(\zeta).
\label{EQ:chirality2}
\end{eqnarray}
In the last equality, we applied the property, $|\zeta|\geq|\eta|$.
A trio of AMCs corresponding to $x_+$ (or $x_-$) share the same chirality. However, the trio of $x_+$'s and the other trio of $x_-$'s have the opposite signs of chiralities.
\item The chiralities of MCs in the ${\rm K'}$-valley are oppposite to those of the corresponding MCs in the ${\rm K}$-valley.
%
\item Majorana quadratic band touching can arise when the trio of $x_+$-AMCs with $\chi(x_+)=-1$ merge into the original ${\rm K}$ point MC with $\chi(x_{\rm K})=+1$. Their merger leads to a Majorana quadratic band touching with the Berry phase of $-2\pi$~\cite{Bena2011, Sticlet2013}. This actually occurs when $\zeta = (1-\eta)/4$.
\item The net chirality in the ${\rm K}$-valley is either $\pm 1$ or $\mp 2$ (Fig.~\ref{fig:AMCs}).
\item The Majorana 
line node is generated in the special case of $\zeta = \eta$ (e.g., by setting $\varGamma^\prime=0$).
In this case, the 
line node
\begin{equation}
~~~~~~
4\eta \cos^2 \frac{\sqrt{3}k_x}{2}  + 4\eta \cos \frac{\sqrt{3}k_x}{2}  \cos \frac{3k_y}{2}+1=4\eta,
\label{EQ:lineNode} 
\end{equation}
appears in place of AMCs for $\eta > \frac{1}{5}$ or $\eta < -\frac{1}{4}$~\cite{Sticlet2013}; 
it forms a closed loop around the $\Gamma$ point for $\eta <-\frac{1}{4}$ or $\eta > \frac{1}{4}$ and around the K and K$^\prime$ points for $\frac{1}{5}<\eta<\frac{1}{4}$, as can be seen from Fig.~\ref{fig:zeta=eta}. 
We note that its 3D analogue has been discussed for the Kitaev model on the hyperhoneycomb lattice \cite{Hermanns2015}.
\end{itemize}

%

%



%



\begin{figure*}
\centering
\includegraphics[width=0.85\linewidth]{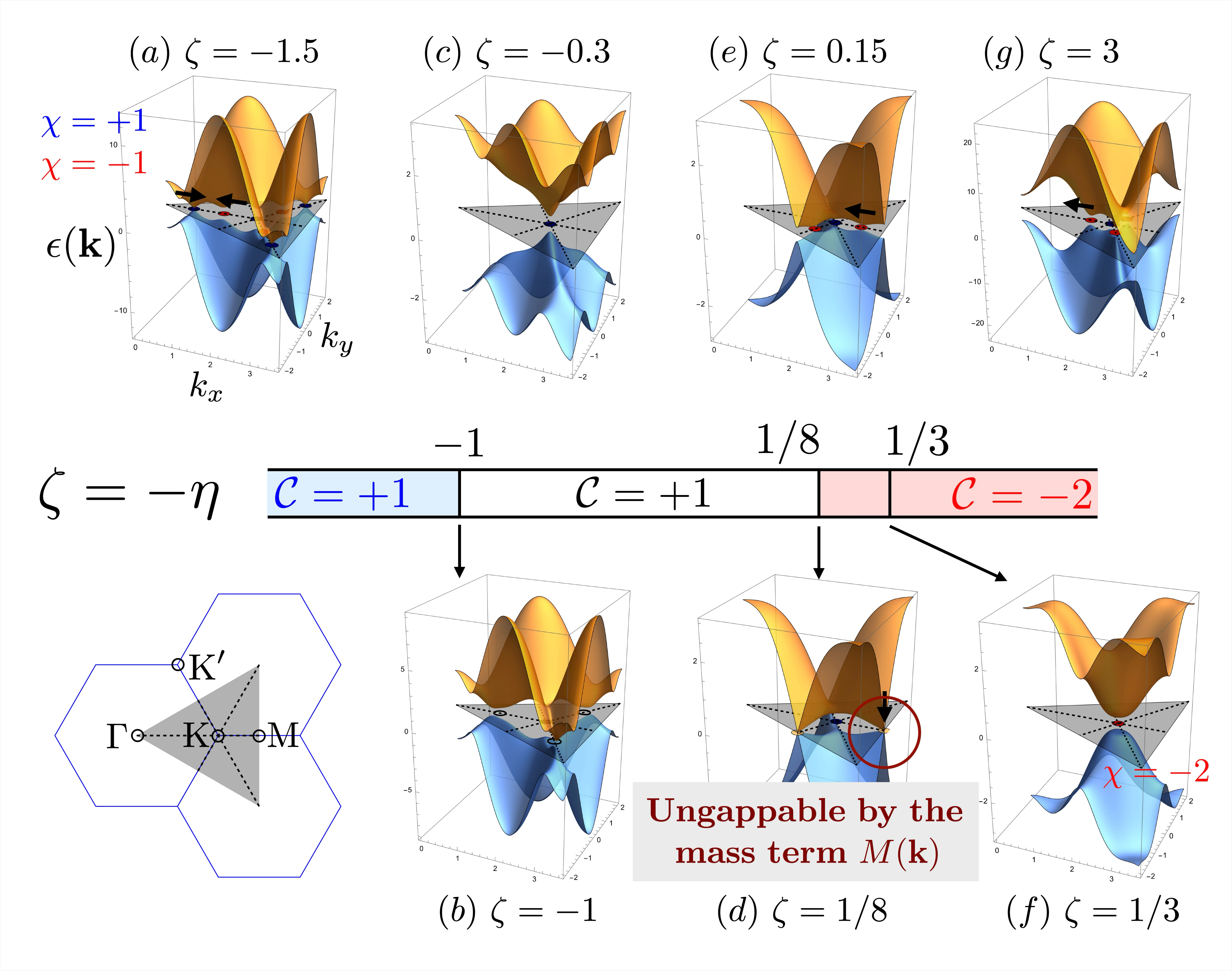}
\caption{
Topological quantum phase transition on the $\zeta=-\eta$ line of Fig.~\ref{fig:chirality} in presence of a magnetic field ($h\ne0$). In the parameter space of $\zeta(=-\eta)$, the system has four distinct states characterized by different Chern numbers and also distinct structures of zero-field Majorana cones: 
(i) $\mathcal{C}=+1$ and $\{x_{\rm K},x_+,x_-\}$ ($\zeta<-1$; blue), 
(ii) $\mathcal{C}=+1$ and $\{x_{\rm K}\}$ ($-1<\zeta<1/8$; white), 
(iii) $\mathcal{C}=-2$ and $\{x_{\rm K},x_+\}$ ($1/8<\zeta<1/3$; red), and 
(iv) $\mathcal{C}=-2$ and $\{x_{\rm K}, x_+\}$ ($1/3<\zeta$; red). 
The energy dispersion $\epsilon({\bf k})$ is illustrated for the four states [(a),(c),(e),(g)] and also at their boundary points [(b),(d),(f)].
In each plot, the energy dispersion is drawn on the gray triangular region centered at the K point of the Brillouin zone (shown in the lower left corner of the figure).
{\bf (a),(c),(e),(g)} In most cases, applied magnetic fields gap out the zero-field Majorana cones existing at $h=0$, whose locations are marked on the Brillouin zone by blue ($\chi=+1$) and red ($\chi=-1$) dots. The black arrows denote the movement directions of the zero-field Majorana cones with increasing $\zeta$.
{\bf (b),(f)} At $\zeta=-1$ and $\zeta=1/3$, the zero-field energy dispersions possess quadratic band touching (QBT), which become gapped out by the applied magnetic field. 
{\bf (d)} At $\zeta=1/8$, the zero-field Majorana cones appearing at three M points remain gapless even under the magnetic field. This critical state connects continuously the two neighboring gapped phases with the different Chern numbers, $\mathcal{C}=+1$ and $\mathcal{C}=-2$, via gap closing at the M points.
}
\label{fig:zeta=-eta}
\end{figure*}

\subsubsection{Topological characteristics for $h \neq 0$}


Associated with the emergence of AMCs for $h=0$ 
is the continuous topological quantum phase transition for $h\neq0$ between the gapped 
non-Abelian KSL with the Chern number $\pm 1$ and the gapped Abelian proximate KSL with the Chern number $\mp 2$ \cite{Kitaev2006, JWang2019, JWang2020, Fujimoto2020, SSZhang2020}. 
Both phases are gapped out by the Zeeman-induced term of Eq.~\eqref{EQ:MFS}:
$M({\bf k})=-\frac{4\varGamma^\prime h}{\sqrt{3}\Delta}\sum_{m=1}^3 \sin \left[{\bf k}\cdot{\bf d}_m^{(2)}\right]$.
Given that the effective Majorana Hamiltonian matrix of Eq.~\eqref{EQ:MajoranaHeff} can now be written as $\boldsymbol{\mathcal F} \cdot \boldsymbol{\sigma}$ where the direction of $\boldsymbol{\mathcal F}=({\rm Re} f, -{\rm Im} f, M)$ is well-defined, the Chern number can be obtained from the familiar formula
$$
\mathcal{C}=\frac{1}{4\pi}\int_{\rm BZ} d^2{\bf k} \boldsymbol{\hat{\mathcal{F}}}\cdot\left(\frac{\partial \boldsymbol{\hat{\mathcal{F}}}}{\partial k_x} \times\frac{\partial \boldsymbol{\hat{\mathcal{F}}}}{\partial k_y}\right).
$$
However, for 
most cases with small $h$, $\mathcal{C}$ 
arises entirely from MCs gapped by $M({\bf k})$, 
in which cases 
the above Chern number formula reduces to the summation over all $i$-th MCs, {\it i.e.}
\begin{equation}
\mathcal{C} = \frac{1}{2}\sum_i \chi_i {\rm sgn} (M_i) ,
\label{EQ:ChernAMC}
\end{equation}
where $\chi_i$ and $M_i$ are the chirality and the mass term $M({\bf k})$ evaluated at the $i$-th MC, respectively~\cite{Sticlet2013}. One key point in evaluating this simpler Chern number formula is that $M({\bf k})$ vanishes only along the $\Gamma{\rm M}$ lines and its sign is constant within a valley but opposite between the two valleys, resulting in $\mathcal{C}\neq0$ from the sets of MCs with opposite chiralities between the two valleys \cite{Haldane1988, Kitaev2006, Chari2021, Hwang2022}.  We have seen from Fig.~\ref{fig:AMCs} that for $h=0$ the AMCs of the same chirality always appear in trios in the same valley, and this can be considered the underlying reason for the emergence of 
the proximate KSL phase with $|\mathcal{C}|>1$ for $h\neq0$. By applying Eq.~\eqref{EQ:ChernAMC} on Fig.~\ref{fig:chirality}, we obtain $\mathcal{C}=\mp 2$ for the red domains and $\mathcal{C}=\pm 1$ elsewhere. Below we will discuss three examples, showing the types and chiralities of the AMCs for $h=0$ along with their locations and the resulting $\mathcal{C}$ when $h\neq0$ gives $M_{\rm K} >0$ (see Fig.~\ref{fig:AMCs} for the illustration). We will also show that the transition between the $\mathcal{C}=\pm 1$ KSL and the $\mathcal{C}=\mp 2$ proximate KSL is continuous in the sense that it is always accompanied by a gap closing 
at the transition point, and that the gap closing occurs either 
at the three ${\rm M}$ points with an 
MC appearing at each (for which $\zeta = \frac{1}{6} + \frac{1}{3}\eta$ or $ac$ line in Fig.~\ref{fig:chirality}) or at the $\Gamma$ point with the 
Majorana quadratic band touching (QBT) (for which $\zeta = -\eta - \frac{1}{2}$ or $ef$ line in Fig.~\ref{fig:chirality}); see Appendix~\ref{APP:QBT3Pi} for details. We emphasize that 
it is $M({\bf k})$ that determines the gap closing point characteristics:
three MCs at the three ${\rm M}$ points or a Majorana QBT at the $\Gamma$ point.

\begin{figure*}
\centering
\includegraphics[width=0.85\linewidth]{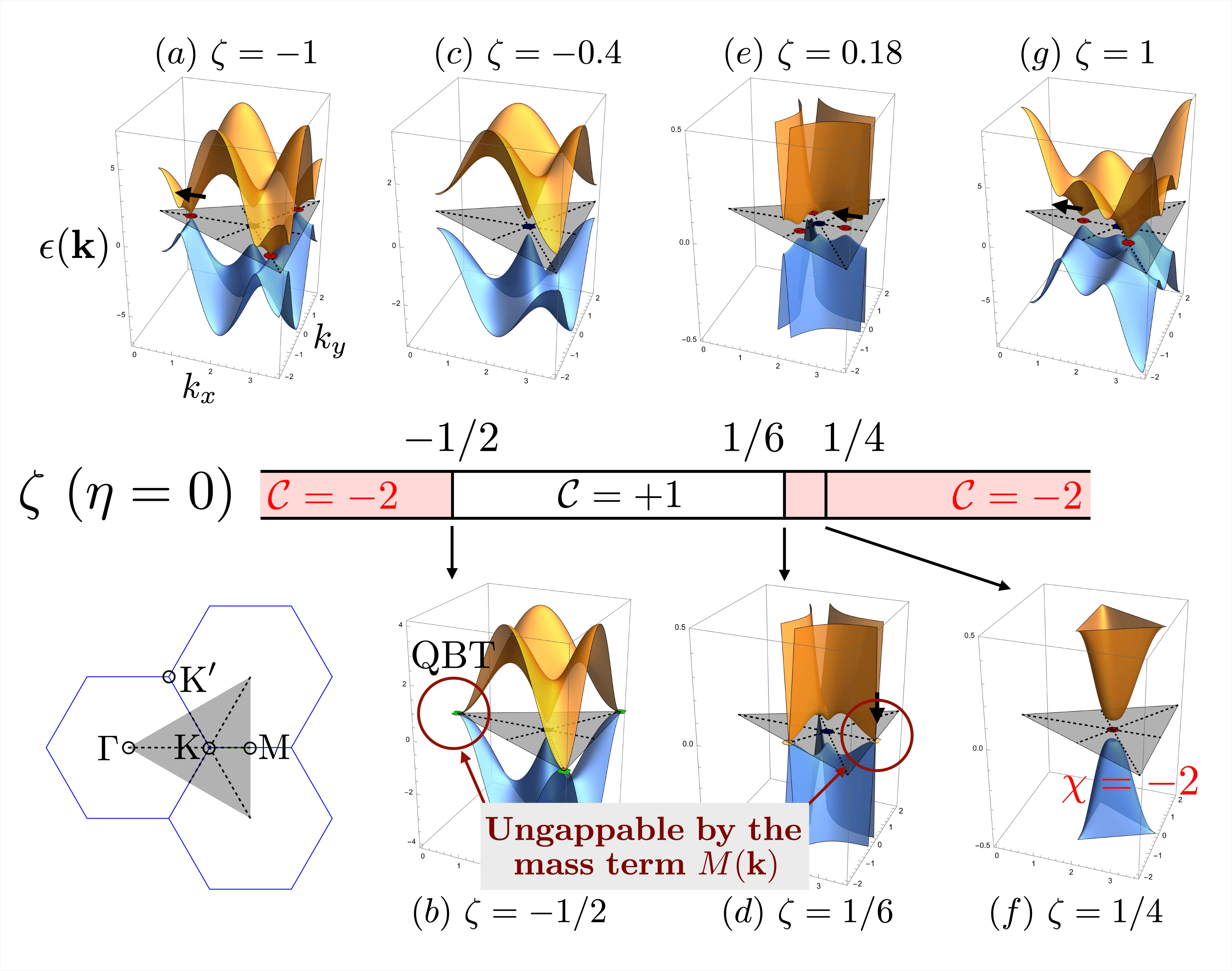}
\caption{
Topological quantum phase transitions on the $\eta=0$ line of Fig.~\ref{fig:chirality} in presence of a magnetic field ($h\ne0$). In the parameter space of $\zeta(\eta=0)$, the system has four distinct states characterized by different Chern numbers and also distinct structures of zero-field Majorana cones: 
(i) $\mathcal{C}=-2$ and $\{x_{\rm K},x\}$ ($\zeta<-1/2$; red), 
(ii) $\mathcal{C}=+1$ and $\{x_{\rm K}\}$ ($-1/2<\zeta<1/6$; white), 
(iii) $\mathcal{C}=-2$ and $\{x_{\rm K},x\}$ ($1/6<\zeta<1/4$; red), and 
(iv) $\mathcal{C}=-2$ and $\{x_{\rm K}, x\}$ ($1/4<\zeta$; red). 
The energy dispersion $\epsilon({\bf k})$ is illustrated for the four states [(a),(c),(e),(g)] and also at their boundary points [(b),(d),(f)].
In each plot, the energy dispersion is drawn on the gray triangular region centered at the K point of the Brillouin zone (shown in the lower left corner of the figure).
{\bf (a),(c),(e),(g)} In most cases, applied magnetic fields gap out the zero-field Majorana cones existing at $h=0$, whose locations are marked on the Brillouin zone by blue ($\chi=+1$) and red ($\chi=-1$) dots. The black arrows denote the movement directions of the zero-field Majorana cones with increasing $\zeta$. 
{\bf (b)} At $\zeta=-1/2$, the zero-field energy dispersion possesses quadratic band touching (QBT) at the $\Gamma$ point, which remain gapless even under the magnetic field. This critical state connects continuously the two neighboring gapped phases with the different Chern numbers, $\mathcal{C}=-2$ and $\mathcal{C}=+1$, via gap closing at the $\Gamma$ point.
{\bf (d)} At $\zeta=1/6$, the zero-field Majorana cones appearing at three M points remain gapless even under the magnetic field. This critical state connects continuously the two neighboring gapped phases with the different Chern numbers, $\mathcal{C}=+1$ and $\mathcal{C}=-2$, via gap closing at the M points.
{\bf (f)} At $\zeta=1/4$, the zero-field energy dispersion possesses QBT at the K point, which become gapped out by the magnetic field.
}
\label{fig:eta=0}
\end{figure*}

\begin{figure*}
\centering
\includegraphics[width=0.85\linewidth]{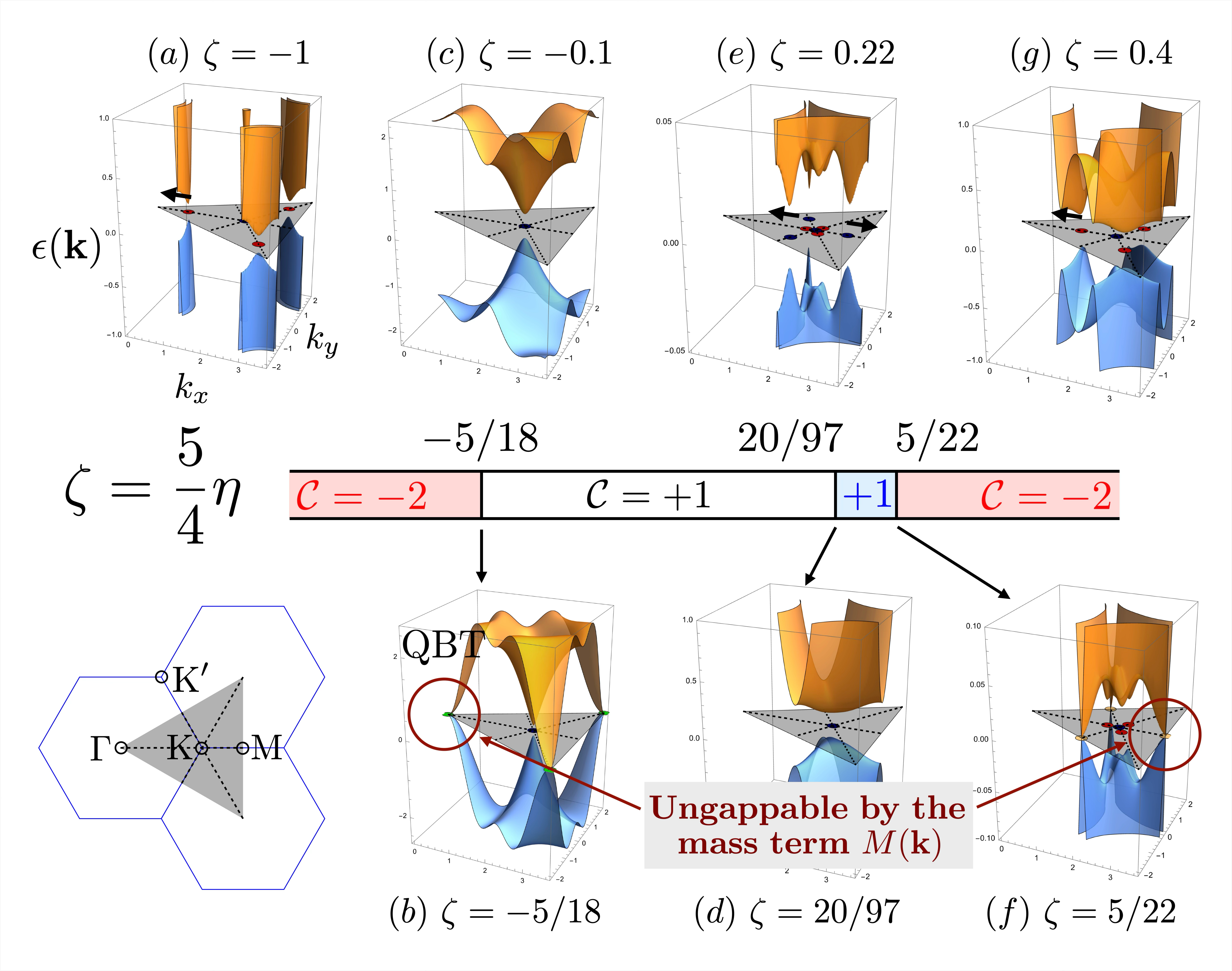}
\caption{Topological quantum phase transitions on the $\zeta=5\eta/4$ line of Fig.~\ref{fig:chirality} in presence of a magnetic field ($h\ne0$).
In the parameter space of $\zeta(=5\eta/4)$, the system has four distinct states characterized by different Chern numbers and also distinct structures of zero-field Majorana cones: 
(i) $\mathcal{C}=-2$ and $\{x_{\rm K},x_-\}$ ($\zeta<-5/18$; red), 
(ii) $\mathcal{C}=+1$ and $\{x_{\rm K}\}$ ($-5/18<\zeta<16/97$; white), 
(iii) $\mathcal{C}=+1$ and $\{x_{\rm K},x_+,x_-\}$ ($16/97<\zeta<5/22$; blue), and 
(iv) $\mathcal{C}=-2$ and $\{x_{\rm K}, x_+\}$ ($5/22<\zeta$; red). 
The energy dispersion $\epsilon({\bf k})$ is illustrated for the four states [(a),(c),(e),(g)] and also at their boundary points [(b),(d),(f)].
In each plot, the energy dispersion is drawn on the gray triangular region centered at the K point of the Brillouin zone (shown in the lower left corner of the figure).
{\bf (a),(c),(e),(g)} In most cases, applied magnetic fields gap out the zero-field Majorana cones existing at $h=0$, whose locations are marked on the Brillouin zone by blue ($\chi=+1$) and red ($\chi=-1$) dots. The black arrows denote the movement directions of the zero-field Majorana cones with increasing $\zeta$. 
{\bf (b)} At $\zeta=-1/2$, the zero-field energy dispersion possesses quadratic band touching (QBT) at the $\Gamma$ point, which remain gapless even under the magnetic field. This critical state connects continuously the two neighboring gapped phases with the different Chern numbers, $\mathcal{C}=-2$ and $\mathcal{C}=+1$, via gap closing at the $\Gamma$ point.
{\bf (d)} At $\zeta=20/97$, the zero-field energy dispersion possesses a single MC at the K point, which become gapped out by the magnetic field.
{\bf (f)} At $\zeta=5/22$, the zero-field Majorana cones located at three M points remain gapless even under the magnetic field. This critical state connects continuously the two neighboring gapped phases with the different Chern numbers, $\mathcal{C}=+1$ and $\mathcal{C}=-2$, via gap closing at the M points.
}
\label{fig:zetaNOT=eta}
\end{figure*}

\begin{figure*}
\centering
\includegraphics[width=0.85\linewidth]{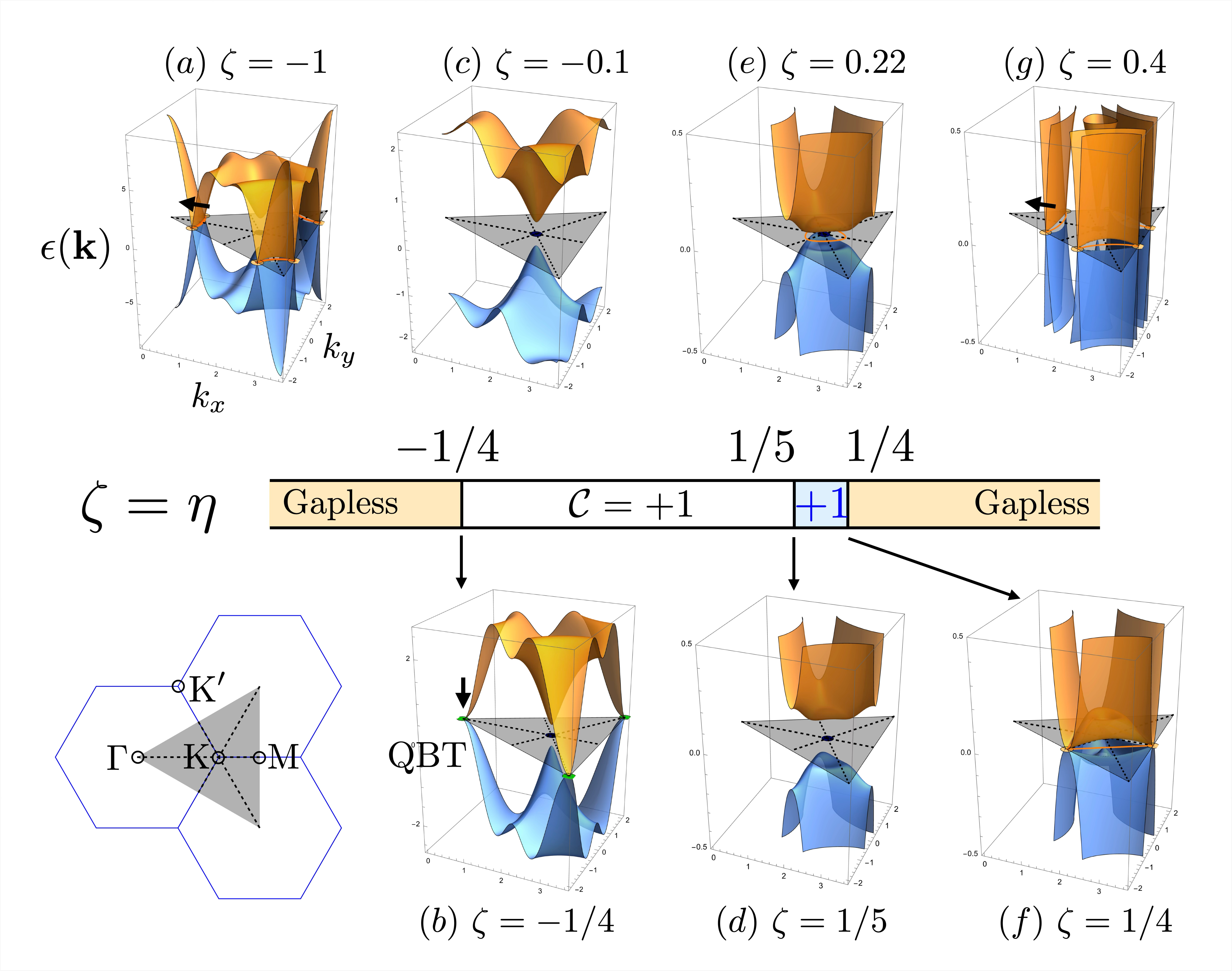}
\caption{Topological quantum phase transitions on the $\zeta=\eta$ line of Fig.~\ref{fig:chirality} in presence of a magnetic field ($h\ne0$).
In the parameter space of $\zeta(=\eta)$, the system has four distinct states: 
(i) gapless ($\zeta<-1/4$; yellow), 
(ii) gapped with $\mathcal{C}=+1$ ($-1/4<\zeta<1/5$; white), 
(iii) gapped with $\mathcal{C}=+1$ ($1/5<\zeta<1/4$; blue), and 
(iv) gapless ($1/4<\zeta$; yellow).
At zero field, except in the region of $-1/4<\zeta<1/5$, Majorana fermi line nodes arise in the energy spectrum as described by Eq.~(\ref{EQ:lineNode}); denoted by orange lines on the Brillouin zone.
Under the magnetic field, the line nodes are mostly gapped out. Only the intersections of the line nodes with the $\Gamma$M lines remain gapless, leading to the gapless states in the yellow regions of the parameter space. In the other regions, the line nodes are all gapped out, resulting in the gapped states with the Chern number $\mathcal{C}=+1$.
The energy dispersion $\epsilon({\bf k})$ is illustrated for the four states [(a),(c),(e),(g)] and also at their boundary points [(b),(d),(f)].
In each plot, the energy dispersion is drawn on the gray triangular region centered at the K point of the Brillouin zone (shown in the lower left corner of the figure).
}
\label{fig:zeta=eta}
\end{figure*}

\begin{itemize}
\item
Our first example is the case with no electric field, {\it i.e.}~$D_1$=0~or~$\zeta = -\eta$.
Fig.~\ref{fig:zeta=-eta} visualizes the topological quantum phase transition occurring in this case.
Details of the change of band topology over four distinct states are discussed below. 
\\\\
\begin{tabular}{ |p{2.5cm}|p{2.5cm}|>{\centering\arraybackslash}p{1cm}|>{\centering\arraybackslash}p{1cm}|}
    \hline
    \multicolumn{4}{|c|}{$D_1 = 0$ limit ($\zeta = -\eta$)} \\
    \hline
    $\zeta$ & AMCs & $\chi$ & $\mathcal{C}$ \\
    \hline
    \multirow{2}{*}{$\zeta\!<\!-1$}
    & $x_+ \in [\Gamma, {\rm K}]$ & $+1$ & \multirow{2}{*}{$+1$}  \\
    \cline{2-3}
    & $x_- \in [\Gamma, {\rm K}]$ & $-1$ &  \\
    \hline
    $-1\!<\!\zeta\!<\!\frac{1}{8}$
    & None &  & $+1$  \\
    \hline
    \multirow{2}{*}{$\frac{1}{8}\!<\!\zeta\!<\!\frac{1}{3}$}
    & $x_+ \in [{\rm M}, {\rm K}]$  & $-1$  & \multirow{2}{*}{$-2$}  \\
    \cline{2-3}
    & $x_- $ doesn't exist & \centering &  \\
    \hline
    \multirow{2}{*}{$ \zeta\!>\!\frac{1}{3}$}
    & $x_+ \in [\Gamma, {\rm K}]$  & $-1$  & \multirow{2}{*}{$-2$}  \\
    \cline{2-3}
    & $x_- $ doesn't exist & \centering &  \\
    \hline
\end{tabular}
\\\\
The energy gap closes only once at $\zeta=\frac{1}{8}$, 
at which the appearance of an  
MC 
at each of the three ${\rm M}$ points signifies a continuous $\Delta\mathcal{C}=3$ topological phase transition. 
Conversely the gap does not close when $\Delta\mathcal{C}=0$ even when there is a change in AMCs for $h=0$. One such case occurs at $\zeta=\frac{1}{3}$ where all AMCs may merge to ${\rm K}$ and ${\rm K'}$ for $h=0$, but this means that $h \neq 0$ gaps out a pair of quadratic band touchings each with a $-2\pi$ Berry phase, maintaining the $\mathcal{C}=-2$ proximate KSL phase. The other case occurs at $\zeta=-1$ where a $x_+$ and $x_-$ pair merge at the midpoint of the $\Gamma{\rm K}$ (or $\Gamma{\rm K'}$) segment for $h=0$, but the opposite chiralities for $x_+$ and $x_-$ AMCs indicate that the $\mathcal{C}=+1$ KSL is maintained.
See Fig.~\ref{fig:zeta=-eta} for the evolution of the energy dispersion.
\item
The second example is the case where the electric field and the hydrostatic pressure have equal effects, {\it i.e.} $D_1 = \varGamma^\prime$ or $\eta=0$; while this is a case where the definition of $x_\pm$ [Eq.~\eqref{EQ:xPM}] is invalid, Eqs.~\eqref{EQ:pRange}, \eqref{EQ:mRange}, and \eqref{EQ:chirality2} still remains valid to the AMCs.
Fig.~\ref{fig:eta=0} visualizes the topological quantum phase transitions occurring in this case.
Details of the change of band topology over four distinct states are discussed below. 
\\\\
\begin{tabular}{ |p{2.5cm}|p{2.5cm}|>{\centering\arraybackslash}p{1cm}|>{\centering\arraybackslash}p{1cm}|}
    \hline
    \multicolumn{4}{|c|}{$D_1 = \varGamma^\prime$ limit ($\eta=0$)} \\
    \hline
    $\zeta$ & AMCs & $\chi$ & $\mathcal{C}$ \\
    \hline
    $\zeta\!<\!-\frac{1}{2}$
    & $x \in [\Gamma, {\rm K}]$ & $-1$ & $-2$  \\
    \hline
    $-\frac{1}{2}\!<\!\zeta\!<\!\frac{1}{6}$
    & None  &  & $+1$  \\
    \hline
    $\frac{1}{6}\!<\!\zeta\!<\!\frac{1}{4}$
    & $x\in [{\rm M}, {\rm K}]$  & $-1$  & $-2$  \\
    \hline
    $ \zeta\!>\!\frac{1}{4}$
    & $x \in [\Gamma, {\rm K}]$  & $-1$  & $-2$  \\
    \hline
\end{tabular}
\\\\
At $\zeta=\frac{1}{6}$, 
the continuous $\Delta\mathcal{C}=3$ topological phase transition is again signified by the appearance of an 
MC at each of the three {\rm M} points. 
On the other hand, at $\zeta = -\frac{1}{2}$, the continuous $\Delta\mathcal{C}=3$ transition signature is the appearance of a 
Majorana QBT 
at the $\Gamma$ point, the details of which is given in Appendix~\ref{APP:QBT3Pi}. 
In this limit, the merging for $h=0$ of a AMC trio with the same chirality with ${\rm K}$ (or ${\rm K}'$) occurs at $\zeta=\frac{1}{4}$; it again leads to a quadratic band touching with $-2\pi$ Berry phase which is gapped out for $h \neq 0$ and maintains the $\mathcal{C}=-2$ proximate KSL.
See Fig.~\ref{fig:eta=0} for the evolution of the energy dispersion.
\item
Lastly, 
we consider the case of an infinitesimally small hydrostatic pressure ($|\varGamma^\prime|\ll|K|$), {\it i.e.} $\eta=(1-\delta)\zeta$ ($0<\delta\ll1$).
Fig.~\ref{fig:zetaNOT=eta} visualizes the topological quantum phase transition occurring in this case.
Details of the change of band topology over four distinct states are discussed below. 
\\\\
\begin{tabular}{ |p{3cm}|p{2.5cm}|>{\centering\arraybackslash}p{0.7cm}|>{\centering\arraybackslash}p{0.7cm}|}
    \hline
    \multicolumn{4}{|c|}{$\varGamma^{\prime} \ll |D_1|$ limit [$\eta=(1-\delta)\zeta$]} \\
    \hline
    $\zeta$ & AMCs & $\chi$ & $\mathcal{C}$ \\
    \hline
    \multirow{2}{*}{$\zeta\!<\!-\frac{1}{4}\!-\!\frac{\delta}{8}$}
    & $x_+ $ doesn't exist & \centering  & \multirow{2}{*}{$-2$}  \\
    \cline{2-3}
    & $x_- \in [\Gamma, {\rm K}]$  & $-1$ &  \\
     \hline
    $-\frac{1}{4}\!-\!\frac{\delta}{8}\!<\!\zeta\!<\!\frac{1}{5}\!+\!\frac{\delta}{25}$
    & None &  & $+1$  \\
    \hline
    \multirow{2}{*}{$\frac{1}{5}\!+\!\frac{\delta}{25}\!<\!\zeta\!<\!\frac{1}{4}\!-\!\frac{\delta}{8}$}
    & $x_+ \in [\Gamma, {\rm K}]$ & $-1$ & \multirow{2}{*}{$+1$}  \\
    \cline{2-3}
    & $x_- \in [{\rm M}, {\rm K}]$ & $+1$ &  \\
    \hline
    \multirow{2}{*}{$\zeta\!>\!\frac{1}{4}\!-\!\frac{\delta}{8}$}
    & $x_+ \in [\Gamma, {\rm K}]$ & $-1$ & \multirow{2}{*}{$-2$}  \\
    \cline{2-3}
    & $x_-$ doesn't exist & \centering &  \\
    \hline
\end{tabular}
\\\\
Once again, the continuous  $\Delta\mathcal{C}=3$ topological phase transition is signified by the appearance of 
an MC 
each of the three ${\rm M}$ points for $\zeta = \frac{1}{4}-\frac{\delta}{8}$ and 
the 
Majorana QBT 
at the $\Gamma$ point for $\zeta = -\frac{1}{4}-\frac{\delta}{8}$. 
However, merging of two AMC trios with the opposite chirality near ${\rm K}$ (or ${\rm K}'$) at $\zeta=\frac{1}{5}+\frac{\delta}{25}$ for $h=0$ leads to a band touching with $0$ Berry phase which is trivially gapped out for $h \neq 0$ and maintains the $\mathcal{C}=+1$ KSL.
See Fig.~\ref{fig:zetaNOT=eta} for the evolution of the energy dispersion for an example where $\delta = \frac{1}{5}$.
\end{itemize}

When the hydrostatic pressure is entirely absent, {\it i.e.} $\varGamma'=0$, the Majorana line node of Eq.~\eqref{EQ:lineNode} is gapped out by the purely Zeeman mass term $\delta M({\bf k})=\frac{72h^3}{\Delta^2}\sum_{m=1}^3 \sin\left[{\bf k}\cdot {\bf d}^{(2)}_m\right]$ which 
adds to $M({\bf k})$ at the level of the third-order perturbation \cite{Kitaev2006}, 
except for the point nodes appearing at a few critical $\zeta$ value, as shown in Fig.~\ref{fig:zeta=eta}; note also that for $\varGamma'\leq 0$ there will be no gap closing in 
$0<|\zeta/\eta-1|\ll 1$, 
as $\frac{\delta M({\bf k})}{M({\bf k})}=-\frac{\varGamma^\prime\Delta}{18\sqrt{3} h^2}>0$ would always hold in that case. 
We note that a 3D analogue of this case has been discussed in 
the hyperhoneycomb lattice 
\cite{Hermanns2015}. 
Point nodes in presence of $\delta M({\bf k})$ 
arise if the Eq.~\eqref{EQ:lineNode} Majorana line node intersects with the $\Gamma{\rm M}$ lines 
where $\delta M({\bf k})$ vanishes. 
As Eq.~\eqref{EQ:lineNode} 
gives us 
no intersection at all for $|\zeta|<\frac{1}{4}$, 
the Chern number $\mathcal{C}=+1$ of the KSL phase is maintained for this range of $\zeta$, as illustrated in Fig.~\ref{fig:zeta=eta}. 
By contrast, for $|\zeta|>\frac{1}{4}$, the Majorana line node of Eq.~\eqref{EQ:lineNode} forms a closed loop around $\Gamma$, thereby giving us a sextet of intersection points 
with the $\Gamma{\rm M}$ line, 
as illustrated in Fig.~\ref{fig:zeta=eta}. This can be attributed to, 
as implied by Eq.~\eqref{EQ:chirality2}, the existence of a $\mathcal{C} = +4$ phase on the other side of $\zeta=\eta$ for $|\zeta|>\frac{1}{4}$ due to the change of the $x_+$ chirality, had $|\zeta|<|\eta|$ been physical.  
Fig.~\ref{fig:zeta=eta} (f) shows that at $\eta = +\frac{1}{4}$, where the Majorana line node goes from encircling ${\rm K}$ to encircling $\Gamma$, the MCs merge in pairs to leave an MC 
at each of three ${\rm M}$ points. Meanwhile at $\eta = -\frac{1}{4}$, Fig.~\ref{fig:zeta=eta} (b) shows that, 
as the Majorana line node 
shrinks and converges into the $\Gamma$ point, all six MCs merge into a quadratic Majorana band touching at $\Gamma$.
%

We note that within the zero flux sector there is a limitation on the possible Chern numbers for the proximate KSL. 
From our perturbation theory in the zero flux sector, either the AMCs exist in a trio for each valley with the chirality opposite of the original MC at ${\rm K}$ or ${\rm K'}$ or the total chiralities of AMCs cancel out, as shown in Fig.~\ref{fig:chirality}. This implies that the continuous quantum phase transition from the KSL to the proximate KSL in our model would always involve the change of the Chern number from $\pm 1$ to $\mp 2$ with the Chern number sign always reversed. This is in contrast with the proximate KSL phases found from the variational Monte Carlo calculations \cite{JWang2019, JWang2020} where discontinuous quantum phase transitions involving the entry of the $\mathbb{Z}_2$ flux entry result in the proximate KSL phases with the Chern number of $\pm 4$, $\pm 5$, and $\mp 8$. In other words, our proximate KSL is always Abelian and therefore the continuous topological phase transition we have obtained is between 
a non-Abelian phase and 
an Abelian phase.

\section{Conclusions and Discussion\label{sec:discussion}}

We have presented a study of the perturbation effect of the DM and the $\varGamma^\prime$ interaction on the zero-flux sector, the main feature of which is a continuous topological phase transition from the non-Abelian $\mathcal{C}=\pm 1$ KSL to the Abelian $\mathcal{C}=\mp 2$ proximate KSL. The key step here is that on the zero flux sector, both the DM and the $\varGamma^\prime$ interaction contribute nothing at the first order but the third and the fourth neighbor Majorana hopping at the second order. When the magnitude of these distant Majorana hopping amplitude grows sufficiently large, the AMCs emerge, which translates into the topological phase transition when $h\neq 0$.  

Our perturbative approach enabled us to scan possibilities of topological phase transitions over all ranges of parameters satisfying $|D_1|,|\varGamma^\prime| \ll |K|$. Given the nature of perturbative approach, 
our chief aim is to complement various numerical analysis, {\it e.g.} the exact numerical calculation results on the effect of 
similar 4-spin interactions on the KSL 
\cite{SSZhang2019, SSZhang2020} 
where the two 4-spin interaction parameters were fixed to have the opposite signs (corresponding to 
$|D_1|>|\varGamma^\prime|$ 
in our model), by examining parameter regions that were left unexplored. Our hope is that this work in turn will serve as a guide for future numerical analysis on the effects of the DM and the $\varGamma^\prime$ interactions that will verify the existence of the continuous topological quantum phase transitions we have discussed.

Tuning electric field and hydrostatic pressure will give us physical access over the $(\eta, \zeta)$ space of Fig.~\ref{fig:chirality}, albeit with a few limitations. From the definition of Eq.~\eqref{EQ:zeta}, we can see that tuning the perpendicular electric field, through $D_1$, will allow us to move in the $(\eta, \zeta)$ space parallel to $\zeta=\eta$, while tuning the hydrostatic pressure, through $\varGamma^\prime$, 
will allow us to move parallel to $\zeta=-\eta$ (see Appendix A and B for details on the effect of electric field and hydrostatic pressure on $D_1$ and $\varGamma^\prime$, respectively, through the modifying the nearest-neighbor electron hopping). However, this does not provide us with any method to change the sign of $\zeta$. 
From Eq.~\eqref{EQ:zeta}, we can expect $\zeta>0$ 
from the antiferromagnetic Kitaev interaction 
$\zeta<0$ 
from the ferromagnetic Kitaev interaction. Fig.~\ref{fig:chirality} suggests that the best bet for accessing our continuous topological phase transition would be to either start from the antiferromagnetic Kitaev interaction or stay near $\zeta=\eta$, {\it i.e.} $|D_1| \gg \varGamma^\prime$.

Transition between the $\mathcal{C}=\pm 1$ KSL and the $\mathcal{C}=\mp 2$ proximate KSL should be manifest in thermal transport measurement. There has been extensive theoretical discussion, with Refs.~\cite{Kitaev2006, Read2000} only being 
most cited examples, on the ratio of thermal Hall conductivity to temperature at low temperatures being proportional to $\mathcal{C}$ [$\kappa_{xy}/T=\mathcal{C}(\pi/12)(k_B^2/\hbar)$], which in more recent years, has also been investigated experimentally \cite{Kasahara2018, Yokoi2021, Imamura2024}. In the context of our results, this implies that the $\mathcal{C}=\pm 1$ to $\mathcal{C}=\mp 2$ transition involves thermal Hall conductivity both reversing its sign and doubling its magnitude.

Additionally, we would like to point out that our {\it continuous} topological phase transition through uniform parameter tuning suggest the possibility of adiabatic deformation and transport of topological 
domains 
through tuning non-uniform parameter. For example, moving  
a finite domain with stronger electric field 
can lead to {\it adiabatic} 
transport of an Abelian $\mathcal{C}=\mp 2$ domain against the non-Abelian $\mathcal{C}=\pm 1$ background with three, {\it odd}, branches of chiral Majorana modes along its boundary. It has been proposed that such 
adiabatic 
deformation of Abelian domains in KSL can be used for fusing a pair of Majorana zero modes \cite{YLiu2022} and 
adiabatically transporting a Majorana zero mode \cite{Klocke2024}. 
Hence, our results suggest that  
topologically protected quantum gates can be realized in the KSL by locally applying electric field or hydrostatic pressure.

\begin{acknowledgments}
We thank Jason Alicea, Minchul Lee, Yuji Matsuda and Pureum Noh for sharing their insights. S.J.K. was supported by the Graduate School of Engineering, The University of Tokyo Doctoral Student Special Incentives Program (SEUT-RA). S.B.C. was supported by the National Research Foundation of Korea (NRF) grants funded by the Korea government (MSIT) (NRF-2023R1A2C1006144 and NRF-2020R1A2C1007554). 
\end{acknowledgments}

\appendix

\section{HOPPING MATRIX AND SPIN INTERACTION PARAMETERS}

In Kitaev materials, the hopping of $d$-orbital electrons create effective spin interactions. \cite{Rau2014, Winter2016} When an electric field is applied to a system, the inversion symmetry of the system is broken by the external field. The breaking of inversion symmetry implies a distortion of the orbitals pushed by the electric field, or the appearance of electric polarization. \cite{SCFuruya2024} Due to this effect on the atomic orbitals, an additional hopping of electrons, which is forbidden by symmetry in the absence of electric fields, emerges as a result of inversion symmetry breaking. For simplicity, we consider the bonding of only two transition atoms, i.e., $Z$-bonding. The hopping matrix $T_Z$ of the $Z$-bonding system, written in the $t_{2g}$ basis without external electric fields, is given by \cite{Rau2014}
\begin{equation}
    T_Z = 
    \begin{pmatrix}
        t_1 & t_2 & 0 \\ t_2 & t_1 & 0 \\ 0 & 0 & t_3
    \end{pmatrix}
\end{equation}

If an electric field is applied, the inversion symmetry-breaking hopping will be introduced. To consider a simple model, we assume that the electric field is applied perpendicular to the sample. To describe the effect of the electric field, we need to adopt two types of coordinate. The first is the $xyz$-coordinate which is consistent with the direction of the $d$-orbitals. The second is the $uvw$-coordinate, defined as \cite{Bolens2018, Chari2021}
\begin{equation}
    \hat{u} = \frac{\hat{x} - \hat{y}}{\sqrt{2}}, \; \hat{v} = \frac{\hat{x} + \hat{y}}{\sqrt{2}}, \; \hat{w} = \hat{z}
\end{equation}

The electric field perpendicular to a sample is given by
\begin{equation}
    \vec{E} = \frac{E}{\sqrt{3}} \hat{x} + \frac{E}{\sqrt{3}} \hat{y} + \frac{E}{\sqrt{3}} \hat{z}
\end{equation}
In the $xyz$-coordinate system, the hopping matrix with inversion symmetry breaking is given by
\begin{equation}
    T_Z = 
    \begin{pmatrix}
        t_1 & t_2 - t_w & t_4 - \sqrt{2} t^{\prime} \\
        t_2 + t_w & t_1 & t_4 + \sqrt{2} t^{\prime} \\
        t_4 + \sqrt{2} t^{\prime} & t_4 - \sqrt{2} t^{\prime} & t_3
    \end{pmatrix}
    \label{EQ:hopping_parameter}
\end{equation}
where $t_u = - t_v = t^{\prime}$ and $t_w$ are the hoppings induced by inversion symmetry breaking, and $t_4$ is the hopping caused by the trigonal distortion.

Following the second order perturbation analysis, we can calculate the interaction coefficients of the spin Hamiltonian in terms of the hopping parameter defined in Eq.~\eqref{EQ:hopping_parameter}. In particular, the DM interaction coefficients are given by \cite{Rau2014, Winter2016}
\begin{equation}
\begin{split}
        D_z^x = D_z^y = & \frac{16 \sqrt{2}}{9} \left[ \mathbb{A} (2t_1 + t_3) +  \mathbb{B}(t_1 + \frac{t_2}{3} - t_3) \right]t^{\prime} \\ & - \frac{16 \mathbb{B}}{3} t_4 t_w
\end{split}
\label{EQ:DM1}
\end{equation}
\begin{equation}
    D_z^z = \frac{16}{9} \left[ -\mathbb{A} (2t_1 + t_3) + 2 \mathbb{B} (t_1 - t_3) \right] t_w
\label{EQ:DM2}
\end{equation}
where $\mathbb{A}$ and $\mathbb{B}$ are coefficients that include Hund’s coupling $J_H$, Coulomb coupling $U$, and spin-orbit coupling $\lambda$, and the subscript $z$ denotes coefficients at $z$-linked system.
\begin{subequations}
\begin{eqnarray*}
\mathbb{A} & = & -\frac{1}{3} \left[ \frac{J_H + 3(U + 3\lambda)}{6J_H^2 - U(U+3\lambda) + J_H(U + 4\lambda)} \right] \\
\mathbb{B} & = & \frac{4}{3} \left[ \frac{3J_H - U - 3\lambda}{6J_H - 2U - 3\lambda} \eta \right] \\
\eta & = & \frac{J_H}{6J_H^2 - J_H(8U + 17\lambda) + (2U + 3\lambda)(U + 3\lambda)}
\end{eqnarray*}
\end{subequations}
Since Eq.~\eqref{EQ:DM1} and Eq.~\eqref{EQ:DM2} implies that the DM interaction is described by two parameters, we define $D_1 = D_z^x = D_z^y$ and $D_2 = D_z^z$. For the system has $C_3$-symmetry, interaction coefficients are not depend on the bonding direction, so we can define $K_{\alpha} = K$, $\varGamma_{\alpha}^{\beta \gamma} = \varGamma$, $\varGamma_{\alpha}^{\alpha \beta} = \varGamma_{\alpha}^{\gamma \alpha} = \varGamma^{\prime}$ and these coefficients also can be obtained by the second order perturbation theory.
\begin{subequations}
    \begin{eqnarray*}
        K & = & \frac{32 \mathbb{A}}{9} (t_w^2 - 2 {t^{\prime}}^2) \\ 
        && + \frac{8 \mathbb{B}}{9} \left[ 3 (t_1 - t_3)^2 - 9 t_2^2 + 9 t_4^2 + 2{t^{\prime}}^2 - t_w^2 \right] \\
        \varGamma & = & \frac{64 \mathbb{A}}{9}{{t^{\prime}}^2} + \frac{8 \mathbb{B}}{9} \left[ 6(t_1 - t_3) t_2 + 9t_4^2 - 2{t^{\prime}}^2 \right] \\
        \varGamma^{\prime} & = & -\frac{32 \sqrt{2} \mathbb{A}}{9} t^{\prime} t_w - \frac{8 \mathbb{B}}{9} \left[ 3(t_1 - t_3) t_4 - 9 t_2 t_4 - \sqrt{2} t^{\prime} t_w \right] \\
        J & = & \frac{4 \mathbb{A}}{9} \left[ (2t_1^2 + t_3)^2 - 4 t_w^2 \right] \\
        && - \frac{8 \mathbb{B}}{9} \left[ 2(t_1 - t_3)^2 + 9 t_4^2 -6{t^{\prime}}^2 - 2t_w^2 \right]
    \end{eqnarray*}
\end{subequations}

\section{INVERSION SYMMETRY BREAKING HOPPING PARAMETERS}
In the $Z$-bonding system, we can define two types of coordinates, the $uvw$-coordinate and the $xyz$-coordinate. The $xyz$-coordinate is consistent with the direction of the $d$-orbitals. At first, we will consider the hopping matrix in the $uvw$-coordinate without the electric field, but with trigonal distortion. \cite{Bolens2018}
\begin{equation}
    T_Z = 
    \begin{pmatrix}
        t_{11} & 0 & 0 \\
        0 & t_{22} & \tilde{t} \\
        0 & \tilde{t} & t_{33}
    \end{pmatrix}_{uvw}
\end{equation}
The subscript $uvw$ indicates that the matrix is expressed in the $uvw$-coordinate system. In the $xyz$-coordinate, the hopping matrix is converted into the well known form as \cite{Rau2014a}
\begin{equation}
    T_Z = 
    \begin{pmatrix}
        \frac{t_{11} + t_{22}}{2} & \frac{-t_{11} + t_{22}}{2} & \frac{\tilde{t}}{\sqrt{2}} \\
        \frac{-t_{11} + t_{22}}{2} & \frac{t_{11} + t_{22}}{2} & \frac{\tilde{t}}{\sqrt{2}} \\
        \frac{\tilde{t}}{\sqrt{2}} & \frac{\tilde{t}}{\sqrt{2}} & t_{33}
    \end{pmatrix}_{xyz}
    =
    \begin{pmatrix}
    t_1 & t_2 & t_4 \\ t_2 & t_1 & t_4 \\ t_4 & t_4 & t_3
    \end{pmatrix}_{xyz}
\label{EQ:hopping matrix}
\end{equation}
where the substitutions $t_1 = \frac{t_{11} + t_{22}}{2}$, $t_2 = \frac{-t_{11} + t_{22}}{2}$, $t_3 = t_{33}$ and $t_4 = \frac{\tilde{t}}{\sqrt{2}}$ are used. 

Next, we consider the hopping induced by inversion symmetry breaking. They will given by \cite{Bolens2018}
\begin{eqnarray}
    T_u & = & \begin{pmatrix} 0 & 0 & -t_u \\ 0 & 0 & -t_u \\ t_u & t_u & 0 \end{pmatrix}_{uvw} \\
    T_v & = & \begin{pmatrix} 0 & 0 & t_v \\ 0 & 0 & -t_v \\ -t_v & t_v & 0 \end{pmatrix}_{uvw} \\
    T_w & = & \begin{pmatrix} 0 & -t_w & 0 \\ t_w & 0 & 0 \\ 0 & 0 & 0 \end{pmatrix}_{uvw}
\end{eqnarray}
Add these equations to Eq.~\eqref{EQ:hopping matrix} and convert them into the $XYZ$-coordinate.
\begin{equation}
    \begin{split}
        T_Z = & \begin{pmatrix}
        t_{11} & -t_w & -t_u + t_v \\
        t_w & t_{22} & \tilde{t} -t_u -t_v \\
        t_u - t_v & \tilde{t} + t_u + t_v & t_{33}
        \end{pmatrix}_{uvw} \\
        = & \begin{pmatrix}
        t_1 & t_2 - t_w & t_4 - \sqrt{2} t_u \\
        t_2 + t_w & t_1 & t_4 - \sqrt{2} t_v \\
        t_4 + \sqrt{2} t_u & t_4 + \sqrt{2} t_w & t_3
        \end{pmatrix}_{xyz}
    \end{split}
\end{equation}
In addition to considering mirror symmetry between the transition atoms, i.e., $t_u = -t_v = t^{\prime}$, the final form of the hopping matrix is given by
\begin{equation}
    T_Z = \begin{pmatrix}
        t_1 & t_2 - t_w & t_4 - \sqrt{2} t^{\prime} \\
        t_2 + t_w & t_1 & t_4 + \sqrt{2} t^{\prime} \\
        t_4 + \sqrt{2} t^{\prime} & t_4 - \sqrt{2} t^{\prime} & t_3
    \end{pmatrix}_{xyz}
\end{equation}

\section{GENERIC PERTURBATION OF SPIN HAMILTONIAN}
In this section, we will show the perturbation approach in section III-(B-E) without ${\rm C}_3$-rotational symmetries. The model Hamiltonian without ${\rm C}_3$ symmetry is given by the combination of the following
\begin{eqnarray*}
    H_K^{\alpha} &=& \sum_{\langle i, j \rangle_\alpha}  K_{\alpha} \sigma_i^{\alpha} \sigma_j^{\alpha} \\
    H_{\varGamma}^{\alpha} &=& \sum_{\langle i, j \rangle_\alpha} \sum_{\beta \neq\alpha} \sum_{\gamma \neq \alpha,\beta}\varGamma_{\alpha}^{\beta \gamma} (\sigma_i^{\beta} \sigma_j^{\gamma} + \sigma_i^{\gamma} \sigma_j^{\beta})/2 \\
    H_{\varGamma^{\prime}}^{\alpha} &=& \sum_{\langle i, j \rangle_\alpha}  \sum_{\beta \neq \alpha} \varGamma_{\alpha}^{\alpha \beta} (\sigma_i^{\alpha} \sigma_j^{\beta} + \sigma_i^{\beta} \sigma_j^{\alpha}) \\
    H_D^{\alpha} &=& \sum_{\langle i, j \rangle_\alpha}  \sum_{\beta,\gamma}\varepsilon_{\alpha\beta\gamma}\left[ 
    D_{\alpha}^{\alpha} \sigma_i^{\beta} \sigma_j^{\gamma} 
    + D_{\alpha}^{\beta} (\sigma_i^{\gamma} \sigma_j^{\alpha} - \sigma_i^{\alpha} \sigma_j^{\gamma}) \right] \\
    H_Z &=& - \sum_i  (h_x \sigma_i^x + h_y \sigma_i^y + h_z \sigma_i^z),
\end{eqnarray*}
of which the first gives us the zeroth order term, {\it i.e.} $H_0 = \sum_\alpha H_K^\alpha$.

The nearest-neighbor Majorana hopping contribution 
from the second-order perturbation in the zero-flux sector by the product of a pair of Zeeman terms 
is given by 
\begin{equation}
        H_{ZZ} = \sum_{\langle A, B \rangle_\alpha} P_0 
        \frac{h_\alpha^2}{\Delta} \sigma_{A}^\alpha \sigma_{B}^\alpha 
         P_0 = 
        -\frac{2 i}{\Delta} \sum_{\langle A, B \rangle_\alpha} h_\alpha^2 c_{A} c_{B}
\end{equation}

\begin{widetext}
The nearest-neighbor Majorana  hopping contribution 
from the second-order perturbation in the zero-flux sector by the product of a pair of spin interactions belonging to the $D_2$-type as defined by the Tab.~\ref{tab:classification-non-Kitaev} is given by
\begin{equation}
    \begin{split}
        H_{D_2 D_2}^\alpha = & \sum_{\langle A,B \rangle_\alpha}\sum_{\beta \neq\alpha} \sum_{\gamma \neq \alpha,\beta}  P_0 \left[ \frac{(\varGamma^{\beta \gamma}_{\alpha} + D_{\alpha}^{\alpha})(\varGamma^{\beta \gamma}_{\alpha} - D_{\alpha}^{\alpha})}{\Delta}  \left( \sigma_{A}^\beta \sigma_{B}^\gamma \sigma_{A}^\gamma \sigma_{B}^\beta + \sigma_{A}^\gamma \sigma_{B}^\beta \sigma_{A}^\beta \sigma_{B}^\gamma \right) \right] P_0 \\
        = & \sum_{\langle A,B \rangle_\alpha}  \frac{2\left[\left({\varGamma_{\alpha}^{\beta \gamma}}\right)^2 - \left({D_\alpha^\alpha}\right)^2\right]}{\Delta} P_0 \sigma_{A}^{\alpha} \sigma_{B}^{\alpha} P_0 \\
        = & \sum_{\langle A, B \rangle_\alpha} \frac{2 i \left[\left({D_{\alpha}^{\alpha}}\right)^2 - \left({\varGamma_{\alpha}^{\beta \gamma}}\right)^2\right]}{\Delta} c_{A} c_{B}
    \end{split}
\end{equation}

The third-neighbor Majorana hopping contribution from the second-order perturbation in the zero-flux sector by the product of a pair of spin interactions belonging to the $D_1/\varGamma^\prime$ type as defined by the Tab.~\ref{tab:classification-non-Kitaev}  
is given by
    \begin{equation}
        \begin{split}
            H_{\varGamma^{\prime} \varGamma^{\prime}}^{(3), \alpha \beta} =
            \sum_{\langle A B A^{\prime} B^{\prime} \rangle_{\alpha \gamma \beta}} & P_0 \left[ 
            \frac{2 (\varGamma^{\alpha \beta}_{\alpha} + D^{\gamma}_{\alpha})(\varGamma^{\alpha \beta}_{\beta} + D^{\gamma}_{\beta})}{\Delta} 
            \sigma_{A}^{\alpha} \sigma_{B}^{\beta} \sigma_{A^{\prime}}^{\alpha} \sigma_{B^{\prime}}^{\beta} 
            + \frac{2 (\varGamma^{\alpha \beta}_{\alpha} - D^{\gamma}_{\alpha})(\varGamma^{\alpha \beta}_{\beta} - D^{\gamma}_{\beta})}{\Delta} 
            \sigma_{A}^{\beta} \sigma_{B}^{\alpha} \sigma_{A^{\prime}}^{\beta} \sigma_{B^{\prime}}^{\alpha} \right] P_0 \\
            = \sum_{\langle A, B \rangle_{3, \alpha \beta}} \sum_{\gamma \neq \alpha, \beta} & 
            \frac{4 i (\varGamma_\alpha^{\alpha \beta} \varGamma_\beta^{\alpha \beta} + D_\alpha^{\gamma} D_\beta^{\gamma})}{\Delta} 
            c_{A} c_{B}
        \end{split}
    \end{equation}
where $\langle A, B \rangle_{3, \alpha \beta}$ implies that the third-neighbor pairs with the combination of the $\alpha$-link and $\beta$-link at the beginning and the ending or vice versa; note 
that the figures both Eqs.~\eqref{EQ:D1p} and ~\eqref{EQ:D1m} would give us $\alpha = x$ and $\beta=y$.

The fourth-neighbor Majorana hopping contribution 
from the second-order perturbation in the zero-flux sector by the product of a pair of spin interactions belonging to the $D_1/\varGamma^\prime$ type by the Tab.~\ref{tab:classification-non-Kitaev} is given as
    \begin{equation}
        \begin{split}
            H_{\varGamma^{\prime} \Gamma^{\prime}}^{(4), \alpha} = &
            \sum_{\langle ABA^\prime B^\prime \rangle_{\alpha\gamma\alpha}} \sum_{\beta \neq \gamma}  
            \frac{2\left\{ \left(\varGamma^{\alpha \beta}_{\alpha}\right)^2 - \left(D^{\gamma}_{\alpha}\right)^2)\right\}}{\Delta} 
            P_0\sigma_{A}^{\alpha} \sigma_{B}^{\beta} \sigma_{A^{\prime}}^{\beta} \sigma_{B^{\prime}}^{\alpha}P_0 
            + \sum_{\langle ABA^\prime B^\prime \rangle_{\alpha\beta\alpha}} \sum_{\gamma \neq \beta}\frac{2\left\{ \left(\Gamma^{\gamma \alpha}_{\alpha}\right)^2 - \left(D^{\beta}_{\alpha}\right)^2)\right\}}{\Delta} 
            P_0 \sigma_{A}^{\gamma} \sigma_{B}^{\alpha} \sigma_{A^{\prime}}^{\alpha} \sigma_{B^{\prime}}^{\gamma} P_0  \\
            = & \sum_{\langle A, B \rangle_{4, \alpha \beta}} 
            \frac{2 i \left[ \left(D_\alpha^\gamma\right)^2 - \left(\varGamma_\alpha^{\alpha \beta}\right)^2 \right]}{\Delta}
            c_{A} c_{B}
            + \sum_{\langle A, B \rangle_{4, \alpha \gamma}} 
            \frac{2 i \left[ \left(D_\alpha^\beta\right)^2 - \left(\varGamma_\alpha^{\gamma \alpha}\right)^2 \right]}{\Delta} 
            c_{A} c_{B}
        \end{split}
    \end{equation}
where $\langle A, B \rangle_{4, \alpha \beta}$ represents a fourth-neighbor pair linked by the initial and the final $\alpha$-links with a $\beta$-link coming in between.

The second-neighbor Majorana hopping contribution 
from the second-order perturbation in the zero-flux sector by the prduct of one Zeeman-type and one $\Gamma$-type spin interactions is given by
    \begin{equation}
        \begin{split}
            H^{(2), \alpha}_{Z\Gamma^{\prime}} = & 
            P_0\left[  
            \sum_{\langle ABA^\prime\rangle_{\alpha\gamma}}\sum_{\beta \neq \alpha,\gamma}\frac{2 (\varGamma^{\alpha \beta}_{\alpha} + D^{\gamma}_{\alpha}) h^{\gamma}}{\Delta}
            \sigma_{A}^{\alpha} \sigma_{B}^{\beta} \sigma_{A^{\prime}}^{\gamma} 
            + \sum_{\langle BAB^\prime \rangle_{\beta\alpha}}\sum_{\gamma \neq \alpha,\beta} \frac{2 (\varGamma^{\gamma \alpha}_{\alpha} + D^{\beta}_{\alpha}) h^{\beta}}{\Delta}
            \sigma_{A}^{\gamma} \sigma_{B}^{\beta} \sigma_{B^{\prime}}^{\alpha}\right. \\
            & \left.
            + \sum_{\langle BAB^\prime\rangle_{\alpha\gamma}}\sum_{\beta \neq \alpha,\gamma}\frac{2 (\varGamma^{\alpha \beta}_{\alpha} - D^{\alpha}_{\gamma}) h^{\gamma}}{\Delta}
            \sigma_{A}^{\beta} \sigma_{B}^{\alpha} \sigma_{B^{\prime}}^{\gamma}  + \sum_{\langle ABA^\prime\rangle_{\beta\alpha}}\sum_{\gamma \neq \alpha,\beta}\frac{2 (\varGamma^{\gamma \alpha}_{\alpha} - D^{\alpha}_{\beta}) h^{\beta}}{\Delta}
            \sigma_{A}^{\beta} \sigma_{B}^{\gamma} \sigma_{A^{\prime}}^{\alpha} \right]P_0
        \end{split}
    \end{equation}
from which we obtain
    \begin{equation}
        \begin{split}
            H^{(2)}_{Z\varGamma^{\prime}} = & 
            \left[ \sum_{\left< A, A^{\prime} \right>_{2, zx}} \frac{2 i \left\{ 
            (\varGamma_x^{xy} + D_x^z) h^z + (\varGamma_z^{yz} - D_z^x)h^x \right\}}{\Delta}
            + \sum_{\left< A, A^{\prime} \right>_{2, yz}} \frac{2 i \left\{ 
            (\varGamma_z^{zx} + D_z^y)h^y + (\varGamma_y^{xy} - D_y^z) h^z \right\}}{\Delta} \right. \\
            & \left. + \sum_{\left< A, A^{\prime} \right>_{2, xy}} \frac{2 i \left\{ 
            (\varGamma_y^{yz} + D_y^x)h^x + (\varGamma_x^{zx} - D_x^y) h^y \right\}}{\Delta} \right] c_{A} c_{A^{\prime}} \\
            & + \left[ \sum_{\left< B, B^{\prime} \right>_{2, zx}} \frac{2 i \left\{ 
            (\varGamma_y^{xy} + D_y^z) h^z + (\varGamma_z^{yz} - D_z^x)h^x \right\}}{\Delta} 
            + \sum_{\left< B, B^{\prime} \right>_{2, yz}} \frac{2 i \left\{ 
            (\varGamma_z^{zx} + D_z^y) h^y + (\varGamma_y^{xy} - D_y^z)h^z \right\}}{\Delta} \right. \\
            & \left. + \sum_{\left< B, B^{\prime} \right>_{2, xy}} \frac{2 i \left\{ 
            (\varGamma_z^{yz} + D_z^x)h^x + (\varGamma_x^{zx} - D_x^y) h^y \right\}}{\Delta} \right] c_{B} c_{B^{\prime}}
        \end{split}
        \label{EQ:second}
    \end{equation}
where $\langle A, A^{\prime} \rangle_{2, \alpha \beta}$ and $\langle B, B^{\prime} \rangle_{2, \alpha \beta}$ imply that the second-neighbor pairs with the combination of the $\alpha$-link and the $\beta$-link term at the beginning and the ending or vice versa; note that the figures both Eqs.~\eqref{EQ:SecA} and ~\eqref{EQ:SecB} would give us $\alpha = x$, $\beta=z$ and $\alpha = z$, $\beta = x$.

\section{EXPLICIT AMC CONDITION AT $h=0$}
\label{APP:pmExplicit}

The reality condition, $x_\pm \in \mathbb{R}$, leads to the constraint, 
\begin{eqnarray}
x_\pm \in \mathbb{R}:~~~~~
\,\left(\left|\eta - \frac{1}{2}\right| \leq \frac{1}{2}\,\,\,{\rm and}\,\,\,|\zeta + \eta|\geq \sqrt{\eta - \eta^2}\right)\,\,\,{\rm or} 
\,\,\,\,\left(\left|\eta - \frac{1}{2}\right| \geq \frac{1}{2}\right).
\label{EQ:realAMC}
\end{eqnarray} 
From the condition, $|x_{\pm}| \leq 1$, we obtain two more constraints, 
\begin{eqnarray}
|x_{+}| \leq 1 : ~~~~~
\,\left[\eta\!\geq 0\,\,\,{\rm and}\,\,\,\left(-2\zeta\!\leq\zeta\!\leq 2\eta\,\,\,{\rm and}\,\,\,\zeta\! \geq -\eta\!-\!\frac{1}{2}\right)\right]
\,{\rm or}\,\,\,\left(\zeta\!\geq \frac{1}{6}\!+\!\frac{1}{3}\eta\right),
\label{EQ:pRange}
\end{eqnarray}
and
\begin{eqnarray}
|x_{-}| \leq 1 : ~~~~~
\,\left[\eta\!\geq 0\,\,\,{\rm and}\,\,\,\left(-2\zeta\!\leq\zeta\!\leq 2\eta\,\,\,{\rm and}\,\,\,\zeta\!\leq \frac{1}{6}\!+\!\frac{1}{3}\eta \right)\right]
\,{\rm or}\,\,\,\left(\zeta\! \leq -\eta\!-\!\frac{1}{2}\right).
\label{EQ:mRange}
\end{eqnarray}

By combining Eqs.~\eqref{EQ:realAMC}, \eqref{EQ:pRange}, \eqref{EQ:mRange}, and \eqref{EQ:chirality2} with $\zeta^2>\eta^2$ from the definition of $\zeta$ and $\eta$ \cite{Sticlet2013}, it is possible to determine the type and the chirality of any AMCs that may exist for given $\zeta$ and $\eta$. We thereby obtain the full, the explicit condition for the existence of the $x_+$ AMCs with negative chirality in the $K$ valley to be
\begin{eqnarray}
&\,&\left(\eta\!\geq\frac{1}{5}\,\,\,{\rm and}\,\,\,\zeta\! > \eta\right)\,\,\,{\rm or}\nonumber\\
&\,&\left[\frac{1}{10}\leq\eta\!<\frac{1}{5}\,\,\,{\rm and}\,\,\,\left(\sqrt{\eta - \eta^2}- \eta\!<\zeta\!<\frac{1}{4}\!-\!\frac{1}{4}\eta\,\,\,{\rm or}\,\,\,\zeta\!>\frac{1}{4}\!-\!\frac{1}{4}\eta\right)\right]\,\,\,{\rm or}\nonumber\\
&\,&\left[-\frac{1}{8}\leq\eta\!<\frac{1}{10}\,\,\,{\rm and}\,\,\,\left(\frac{1}{6}\!+\!\frac{1}{3}\eta\!<\zeta\!<\frac{1}{4}\!-\!\frac{1}{4}\eta\,\,\,{\rm or}\,\,\,\zeta\!>\frac{1}{4}\!-\!\frac{1}{4}\eta\right)\right]\,\,\,{\rm or}\nonumber\\
&\,&\left[-\frac{1}{3}\leq\eta\!<-\frac{1}{8}\,\,\,{\rm and}\,\,\,\left(-\eta\!\leq\zeta\!<\frac{1}{4}\!-\!\frac{1}{4}\eta\,\,\,{\rm or}\,\,\,\zeta\!>\frac{1}{4}\!-\!\frac{1}{4}\eta\right)\right]\,\,\,{\rm or}\nonumber\\
&\,&\left(\eta<-\frac{1}{3}\,\,\,{\rm and}\,\,\,\zeta\!\geq-\eta\right),
\end{eqnarray}
that of the existence of the $x_+$ AMCs with positive chirality in the $K$ valley to be
\begin{eqnarray}
&\,&\left(\eta\! > 1\,\,\,{\rm and}\,\,\,-\eta\!-\!\frac{1}{2}<\zeta\!\leq-\eta\right)\,\,\,{\rm or}\nonumber\\
&\,&\left(\frac{1}{2}<\eta\!\leq 1\,\,\,{\rm and}\,\,\,-\eta\!-\!\frac{1}{2}<\zeta\!<-\sqrt{\eta - \eta^2}-\eta\right),
\end{eqnarray}
that of the existence of the $x_-$ AMCs with positive chirality in the $K$ valley to be
\begin{eqnarray}
&\,&\left(\frac{1}{5}\leq\eta\!\leq\frac{1}{4}\,\,\,{\rm and}\,\,\,\eta\!<\zeta\! < \frac{1}{6}\!+\!\frac{1}{3}\eta\right)\,\,\,{\rm or}\nonumber\\
&\,&\left[\frac{1}{7}<\eta\!<\frac{1}{5}\,\,\,{\rm and}\,\,\,\left(\sqrt{\eta\!-\eta^2}-\eta\!<\zeta\! < \frac{1}{4}\!-\!\frac{1}{4}\eta\,\,\,{\rm or}\,\,\,\frac{1}{4}\!-\!\frac{1}{4}\eta\!<\zeta\!<\frac{1}{6}\!+\!\frac{1}{3}\eta\right)\right]\,\,\,{\rm or}\nonumber\\
&\,&\left(\frac{1}{10}<\eta\!\leq\frac{1}{7}\,\,\,{\rm and}\,\,\,\sqrt{\eta\!-\eta^2}-\eta\!<\zeta\! < \frac{1}{6}\!+\!\frac{1}{3}\eta\right).
\end{eqnarray}
and that of the existence of the $x_-$ AMCs with negative chirality in the $K$ valley to be
\begin{eqnarray}
&\,&\left(\eta\! > 1\,\,\,{\rm and}\,\,\,\zeta\!\leq-\eta\right)\,\,\,{\rm or}\nonumber\\
&\,&\left(\frac{1}{2}<\eta\!\leq 1\,\,\,{\rm and}\,\,\,\zeta\!<-\sqrt{\eta - \eta^2}-\eta\right)\,\,\,{\rm or}\nonumber\\
&\,&\left(-\frac{1}{4}<\eta\leq\frac{1}{2}\,\,\,{\rm and}\,\,\,\zeta\!<-\eta\!-\!\frac{1}{2}\right)\,\,\,{\rm or}\nonumber\\
&\,&\left(\eta\!\leq-\frac{1}{4}\,\,\,{\rm and}\,\,\,\zeta\!<\eta\right).
\end{eqnarray}
\end{widetext}

\section{GAP CLOSING AT TOPOLOGICAL QUANTUM PHASE TRANSITIONS}
\label{APP:QBT3Pi}

We examine here the energy gap closings 
occurring at M for $\zeta = \frac{1}{6}+\frac{1}{3}\eta$ and the other occurring at $\Gamma$ for $\zeta = -\frac{1}{2}-\eta$. 
This can be most easily done by expanding
\begin{equation}
{\bf \mathcal{F}}({\bf k})\cdot \boldsymbol{\sigma}=\left(\begin{array}{cc}M({\bf k}) & f({\bf k}) \\ f^*({\bf k}) & -M({\bf k})\end{array}\right)
\end{equation}
around M for $\zeta = \frac{1}{6}+\frac{1}{3}\eta$ and around $\Gamma$ for $\zeta=-\frac{1}{2}-\eta$. 

To consider ${\bf \mathcal{F}}({\bf k})$ around ${\rm M}$ for $\zeta = \frac{1}{6}+\frac{1}{3}\eta$, we can take ${\bf k} = \hat{\bf x}2\sqrt{3}\pi + \delta {\bf q}$ and find up to the first order
\begin{eqnarray*}
    f_1({\bf k})&=&-iK_1 \sum_{m=1}^3 e^{-i{\bf k}\cdot{\bf d}_m^{(1)}}\nonumber\\
    &=&-iK_1 \sum_{m=1}^3 e^{-i(\hat{\bf x}2\sqrt{3}\pi + \delta {\bf q})\cdot{\bf d}_m^{(1)}}\nonumber\\ 
    &=& iK_1 + 2K_1\delta q \sin \phi_{\rm M} + O(\delta q^2)
\end{eqnarray*}
(where $\tan \phi_{\rm M} \equiv \delta q_y/\delta q_x$),
\begin{eqnarray*}
    f_3({\bf k})&=&-i2\zeta K_1 \sum_{m=1}^3 e^{-i{\bf k}\cdot{\bf d}_m^{(3)}}= -i2\zeta K_1 \sum_{m=1}^3 e^{i2{\bf k}\cdot{\bf d}_m^{(1)}}\nonumber\\
    &=&-i2\left(\frac{1}{6}+\frac{1}{3}\eta\right)K_1 \sum_{m=1}^3 e^{i2(\hat{\bf x}2\sqrt{3}\pi + \delta {\bf q})\cdot{\bf d}_m^{(1)}}\nonumber\\
    &=&-i(1+2\eta)K_1 + O(\delta q^2),
\end{eqnarray*}
and
\begin{widetext}
 \begin{eqnarray*}
    f_4({\bf k})&=&-i\eta K_1 \sum_{m=1}^6 e^{-i{\bf k}\cdot{\bf d}_m^{(4)}}\nonumber\\
    &=&-i\eta K_1 \left[\left\{\sum_{m=1}^3 e^{-i{\bf k}\cdot{\bf d}_m^{(1)}}\right\}\left\{\sum_{m=1}^3 e^{-i2{\bf k}\cdot{\bf d}_m^{(1)}}\right\}-\sum_{m=1}^3 e^{-i3{\bf k}\cdot{\bf d}_m^{(1)}}\right]\nonumber\\
    &=&-i\eta K_1 \left[\left\{\sum_{m=1}^3 e^{-i(\hat{\bf x}2\sqrt{3}\pi + \delta {\bf q})\cdot{\bf d}_m^{(1)}}\right\}\left\{\sum_{m=1}^3 e^{-i2(\hat{\bf x}2\sqrt{3}\pi + \delta {\bf q})\cdot{\bf d}_m^{(1)}}\right\}-\sum_{m=1}^3 e^{-i3(\hat{\bf x}2\sqrt{3}\pi + \delta {\bf q})\cdot{\bf d}_m^{(1)}}\right]\nonumber\\
    &=&i2\eta K_1 + O(\delta q^2)
\end{eqnarray*}   
\end{widetext}
put together gives
\begin{eqnarray*}
    f({\bf k})&=&f_1({\bf k}) + f_3 ({\bf k})+f_4({\bf k})\nonumber\\
    &=& 2K_1\delta q \sin \phi_{\rm M} + O(\delta q^2),
\end{eqnarray*}
while
\begin{eqnarray*}
    M({\bf k})&=&-K_2\sum_{m=1}^3 \sin \left[(\hat{\bf x}2\sqrt{3}\pi + \delta {\bf q})\cdot{\bf d}_m^{(2)}\right]\nonumber\\
    &=& -2\sqrt{3}K_2 \delta q \cos \phi_{\rm M} + O(\delta q^2).
\end{eqnarray*}
Combined, we obtain around M
\begin{equation}
{\bf \mathcal{F}}({\bf k})\cdot \boldsymbol{\sigma}=2\delta q(-\sigma^z \sqrt{3}K_2 \cos \phi_{\rm M} + \sigma^x 2K_1\sin \phi_{\rm M})
\end{equation}
up to the first order. 
This shows how a Majorana cone can arise with the inclusion of the $M({\bf k})$ contribution. 
There being three ${\rm M}$ points in the 1st BZ, we can conclude that three Majorana cones appear at $\zeta=\frac{1}{6}+\frac{1}{3}\eta$. 

For ${\bf \mathcal{F}}({\bf k})$ around $\Gamma$ for $\zeta = -\frac{1}{2}-\eta + \delta$ up to the third order in $k$ and $\delta \ll k^3$,
\begin{equation*}
    f_1({\bf k})= K_1\left(-i3 +i\frac{3}{4}k^2+ \frac{1}{8} k^3 \sin 3\phi\right)
\end{equation*}
(where $\tan \phi \equiv k_y/k_x$),
\begin{equation*}
    f_3({\bf k})=(1+2\eta-2\delta)K_1(i 3 -i3  k^2 +k^3\sin 3\phi),
\end{equation*}
and
\begin{equation*}
    f_4({\bf k})=\eta K_1 \left(-i 6 + i\frac{9}{2}k^2\right)
\end{equation*}
put together gives
\begin{eqnarray*}
    f({\bf k})&=&f_1({\bf k}) + f_3 ({\bf k})+f_4({\bf k})\nonumber\\
    &=& -iK_1\left[6\delta+\left(\frac{9}{4}+\frac{3}{2}\eta\right) k^2\right]\nonumber\\
    &\,&+\left(\frac{9}{8}+2\eta\right)K_1 k^3 \sin 3\phi,
\end{eqnarray*}
while
\begin{eqnarray*}
    M({\bf k})&=&-K_2\sum_{m=1}^3 \sin \left[{\bf k}\cdot{\bf d}_m^{(2)}\right]\nonumber\\
    &=& \frac{3\sqrt{3}}{2}K_2 k^3 \cos 3\phi.
\end{eqnarray*}
Combined, we obtain around $\Gamma$
\begin{eqnarray}
{\bf \mathcal{F}}({\bf k})\cdot \boldsymbol{\sigma}&=&\frac{3\sqrt{3}}{8}k^3\left[\sigma^z 4K_2 \cos 3\phi + \sigma^x \sqrt{3}\left(1+\frac{16}{9}\eta\right)K_1\sin 3\phi\right]\nonumber\\ 
&\,&+\sigma^y K_1\left[6\delta+\frac{9}{4}\left(1+\frac{2}{3}\eta\right) k^2\right] + O(k^4).
\end{eqnarray}
This shows that, while we have a quadratic band touching at $\Gamma$ for $\zeta = -\frac{1}{2}-\eta$, {\it i.e.} $\delta =0$, 
there occurs a $\Delta \mathcal{C}=3$ topological quantum phase transition as we cross $\zeta = -\frac{1}{2}-\eta$, {\it i.e.} $\delta$ changes its sign.

\bibliography{higherChern}

\end{document}